\documentclass[letterpaper,8pt]{extarticle}  
\usepackage{import}
\usepackage{local}
\usepackage[left=.65in,top=.65in,bottom=.65in,right=.65in]{geometry}
\usepackage{lipsum} 
\usepackage{algorithm}
\usepackage{algpseudocode}
\usepackage{amssymb}
\usepackage[most]{tcolorbox}
\usepackage{booktabs,tabularx,array}
\usepackage{seqsplit}

\newtcolorbox{promptbox}[1]{%
  enhanced, breakable,
  colback=white,            
  colframe=black,           
  boxrule=0.7pt,            
  arc=2mm,                  
  title={#1},               
  fonttitle=\bfseries\color{white},
  colbacktitle=black, coltitle=white, 
  boxed title style={sharp corners, colback=black, colframe=black},
  varwidth boxed title*=-2mm 
}

\providecommand{\tracegap}{\par\smallskip\centerline{\textit{... excerpt omitted ...}}\smallskip}
\providecommand{\traceomittedgenes}{\par\smallskip\textit{Additional lower-priority genes omitted for brevity.}\smallskip}

\title{LA-MARRVEL: A Knowledge-Grounded, Language-Aware LLM Framework for Clinically Robust Rare Disease Gene Prioritization}

\author{%
\textbf{Jaeyeon Lee,\textcolor{Accent}{\textsuperscript{1,2}} %
Lin Yao,\textcolor{Accent}{\textsuperscript{1,2,3}} %
Hyun-Hwan Jeong,\textcolor{Accent}{\textsuperscript{1,2*}} %
Zhandong Liu\textcolor{Accent}{\textsuperscript{1,2,4,*}} }\\
\begin{small}\textcolor{Accent}{\textsuperscript{1}}Department of Pediatrics, Baylor College of Medicine, Houston, Texas, 77030\\ 
\textcolor{Accent}{\textsuperscript{2}}Jan and Dan Duncan Neurological Research Institute, Texas Children's Hospital, Houston, Texas, 77030\\ 
\textcolor{Accent}{\textsuperscript{3}}Department of Child Neurology and Developmental Neurosciences, Texas Children's Hospital, Houston, Texas, 77030\\
\textcolor{Accent}{\textsuperscript{4}}Quantitative and Computational Biosciences program, Baylor College of Medicine, Houston, Texas, 77030\\
\textcolor{Accent}{\textsuperscript{*}}Co-corresponding authors: \textcolor{Accent}{hyun-hwan.jeong@bcm.edu}, \textcolor{Accent}{zhandonl@bcm.edu} \\ \end{small}
}

\date{}

\begin{document}
\maketitle
\thispagestyle{empty}

\begin{doublespacing}
\section{Abstract}


\noindent

\textbf{\textcolor{Accent}{Rare disease diagnosis requires matching variant-bearing genes to complex patient phenotypes across large and heterogeneous evidence sources. 
This process remains time-intensive in current clinical interpretation pipelines.
To overcome these limitations,
We present LA-MARRVEL, a knowledge-grounded, language-aware LLM framework and designed for clinical robustness and practical deployment. LA-MARRVEL delivers a 12--15 percentage-point absolute improvement in Recall@1 over established gene prioritization approaches, showing that architectural design can drive substantial accuracy gains. 
We found that the central contributor is structured, phenotype-rich prompt construction that explicitly encodes patient and disease phenotypes, preserving clinically meaningful context more effectively than disease labels alone.
Across three real-world cohorts, LA-MARRVEL consistently improves gene-ranking performance, including in challenging cases where the causal gene was initially ranked lower by first-stage prioritization. 
For each candidate gene, the system delivers clinically relevant, ACMG-aligned reasoning that integrates phenotype concordance, inheritance patterns, and variant-level evidence into auditable explanations, enabling streamlined clinical review.
These findings suggest that knowledge-grounded LLM layer can enhance existing rare-disease gene prioritization workflows without altering established diagnostic pipelines.}}

\section{Introduction}

Rare diseases collectively affect an estimated 263--446 million individuals worldwide (3.5--5.9\% of the global population)\citep{NguengangWakap2020}. Yet each individual condition requires a complex and burdensome diagnostic process\citep{BenitoLozano2022}. The average time to diagnosis ranges from 4.7 to 7.6 years depending on geography and patient population\citep{Faye2024,Wise2019}. Modern DNA and RNA sequencing tests can identify genes potentially associated with a patient's symptoms, but they typically return extensive candidate lists\citep{IOM1994GeneticRisks, Murdock2020}. Clinicians must then evaluate these candidates against clinical guidelines, multiple databases, and a large and continuously expanding biomedical literature to determine which variant best explains the phenotype. This interpretative step is labor-intensive, time-consuming, and prone to missed connections, particularly when key evidence is embedded in unstructured natural language.

Large language models (LLMs) can read and explain complex medical texts. For example, the recent models show higher than 95\% accuracy for MedQA, a USMLE (United States Medical Licensing Examination)-style benchmark \citep{Jin2021,ValsMedQA2025}. However, our prior work demonstrated that LLM outputs are biased toward well-studied genes and sensitive to candidate gene ordering, compromising the robustness and reliability of gene prioritization
 \cite{Neeley2024}. These limitations suggest that, for clinical decision-support tasks, LLMs must link outputs to citable knowledge (grounding) and produce consistent responses across repeated runs (stability). Without these properties, a strong task-performance alone does not ensure safe clinical use \citep{Wu2025, Asgari2025}.
A major reason for this challenge is that rare disease evidence is heterogeneous. In practice, clinicians must synthesize patient annotation (medical history, symptom evolution, physical examination, diagnostic studies), disease annotation (HPO phenotypes, mode of inheritance, known associations), family information (segregation, de novo status, haplotype phasing), gene annotation (constraint and function metrics), and variant annotation (variant consequence type, allele frequency, ClinVar assertions, computational predictors). This multi-source synthesis is consistent with ACMG/AMP variant-interpretation guidance, which combines molecular evidence with phenotype specificity, segregation, and de novo context for clinical classification \citep{Richards2015}. A substantial proportion of clinical evidence remains in unstructured or semi-structured form \citep{Murdoch2013}. Even when symptoms and inheritance are encoded with ontologies such as HPO, which is structured as a directed acyclic graph, converting heterogeneous sources into fixed machine-learning features still requires substantial normalization and feature engineering. This process may also lead to loss of clinically relevant contextual information \citep{Khler2020,Garla2012}.
Computational gene-prioritization tools for rare disease diagnosis are typically used as a high-recall first stage to filter/rank an initial list of candidate genes, identifying many plausible targets but retaining some a few false positives\citep{Jacobsen2022}. The challenge is integrating these heterogeneous inputs into one coherent ranking signal; this synthesis is error-prone and labor-intensive and diverts valuable time and attention from busy clinical care\citep{Brownstein2014,Dewey2014}. Recent rare-disease AI systems have begun exploring knowledge-guided few-shot phenotype-driven diagnosis and multi-agent conversational LLM reasoning \citep{Alsentzer2025SHEPHERD,Chen2025MAC}. We propose that this stage is where an LLM-based layer provides the greatest value, integrating heterogeneous evidence in language space after first-stage retrieval has narrowed the candidate-set and enriched gene-level annotations.

To bridge these gaps, we introduce LA-MARRVEL, a language-aware reranking system that sits on top of a knowledge-driven high-recall pipeline, AI-MARRVEL\citep{Mao2024}. Rather than supplanting established bioinformatics tools, our goal is to further prioritize candidate genes, yielding a more concise and accurate list. LA-MARRVEL has two core components. First, we provide expert-engineered context: curated phenotype and disease information that gives the model the right facts at the right time. Second, we use a ranked voting method (Tideman's method)\citep{Tideman1987} to combine multiple responses into a single, consensus gene order. This voting step increases ranking stability by aggregating across responses and favoring candidates supported by consistent evidence.

\begin{figure}
\centering
\includegraphics[width=.85\textwidth]{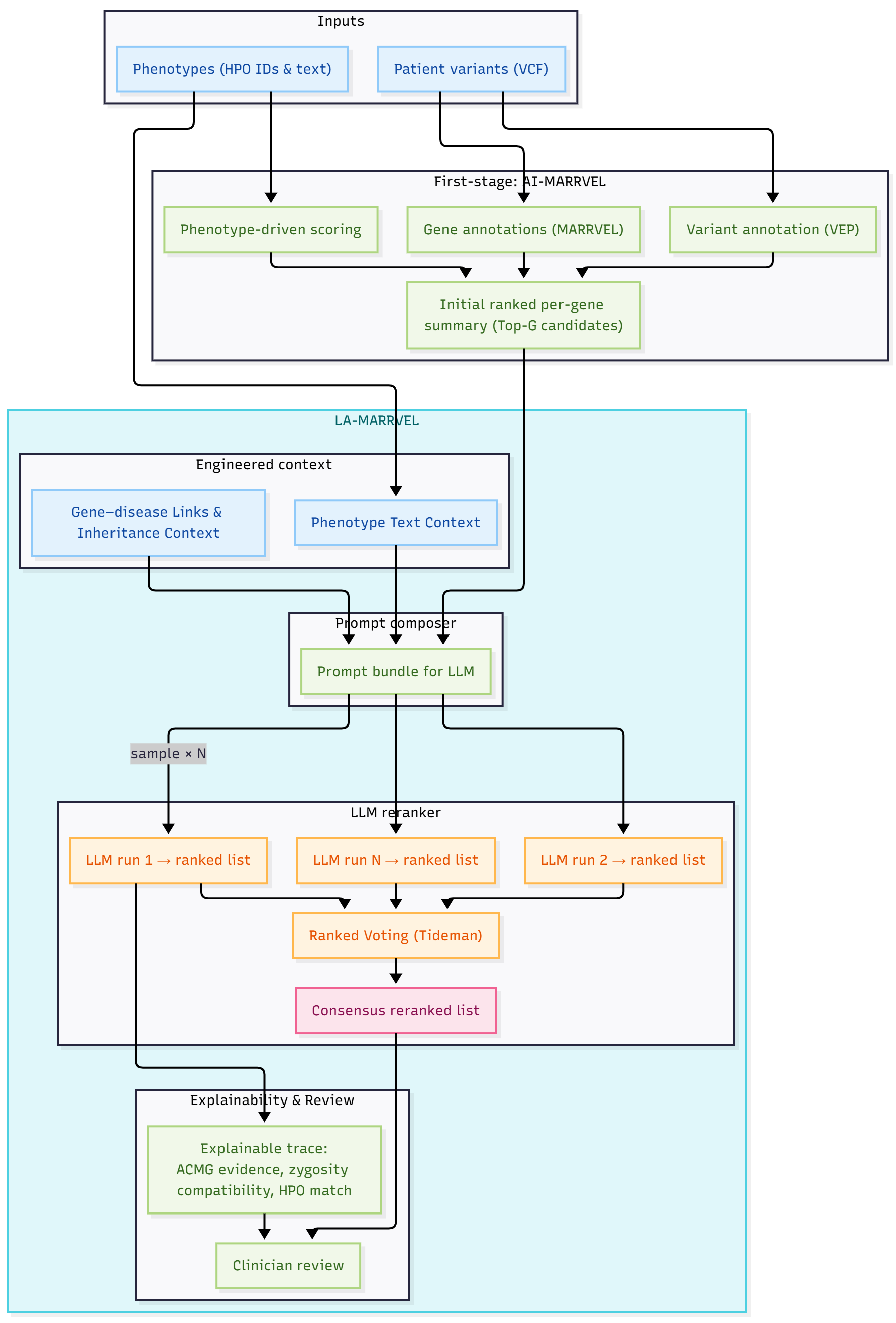}
\caption{\textcolor{Highlight}{
\textbf{Schematic illustration of LA-MARRVEL}} \\
AI-MARRVEL first generates high-recall, variant-bearing candidate genes with annotations. LA-MARRVEL then composes knowledge-grounded prompts using HPO terms, disease and gene summaries, and variant-level evidence, queries an LLM multiple times, and aggregates the resulting partial rankings using Tideman's ranked-pairs voting. The final output is a reranked gene list with an explainable trace that integrates phenotype match, inheritance, and ACMG-style variant assessment.
}
\label{fig:F1}
\end{figure}

We evaluate LA-MARRVEL on three real-world patient cohorts \textemdash Baylor Genetics (BG), Deciphering Developmental Disorders (DDD), and the Undiagnosed Diseases Network (UDN). These datasets come from independent cohorts which reflect real-world diagnosed cohorts from independent sources. We compare our approach to widely used diagnostic tools (e.g., Exomiser\citep{Jacobsen2022} and LIRICAL\citep{Robinson2020}) and to general-purpose LLMs. Across the three cohorts, we demonstrated LA-MARRVEL improves top-rank accuracy while maintaining high recall overall, resulting in more reliable prioritization of the causal gene and consequently reducing the downstream review burden  (\Cref{fig:F1}).


A notable advantage of LA-MARRVEL is its transparency and explainability. For each ranked gene, the system provides plain-language explanations describing how the patient’s phenotypes correspond to known disease features and how observed variants align with the expected mode of inheritance. These explanations are grounded in curated domain knowledge and generated using a consensus-based methodology, enabling users to trace how heterogeneous evidence contributes to the final gene ranking and to critically evaluate the reasoning underlying the result. We therefore present LA-MARRVEL as a language-aware reranking framework designed to improve the robustness and interpretability of rare-disease gene prioritization within existing diagnostic workflows.

\section{Results}

\subsection{Two-Stage Design Improves Retrieval Performance}
\begin{figure}[htbp]
  \captionsetup[subfigure]{labelformat=parens,labelfont={bf,large}} 
  \centering

  \begin{subfigure}{.85\linewidth}
    \caption{ \large Comparison with Naive LLMs by Recall@$K$}
    \centering
    \includegraphics[width=\linewidth]{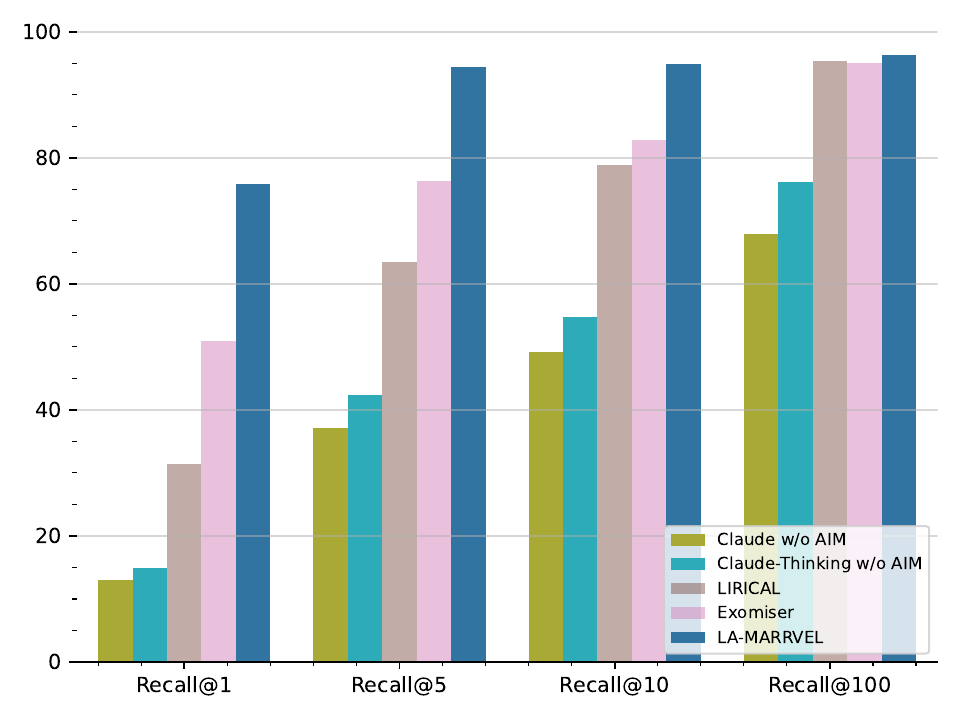}
    \label{fig:F2A}
  \end{subfigure}


  \begin{subfigure}{.85\linewidth}
    \caption{ \large Cross-Dataset Evaluation with Bioinformatics Pipelines}

    \centering
    \includegraphics[width=\linewidth]{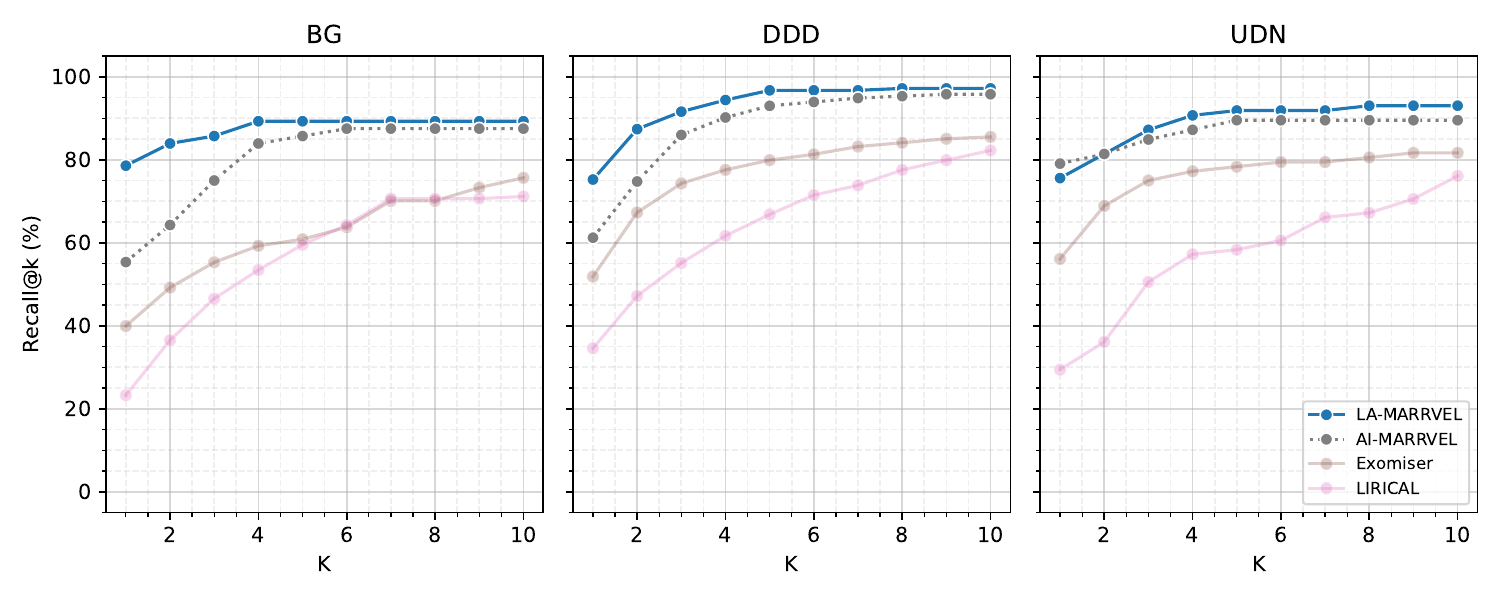}
    \label{fig:F2B}
  \end{subfigure}

  \caption{\textcolor{Highlight}{
    \textbf{Comparative Performance Analysis of LA-MARRVEL.}} \\
    \textbf{(A)} Barplot of Recall@$K$ (K = 1, 5, 10, 100) comparing two prompt-only LLM settings (Claude and Claude-Thinking, both without AI-MARRVEL context) against classical phenotype-driven tools (LIRICAL, Exomiser) and LA-MARRVEL. Prompt-only LLMs recover only \textasciitilde12--15\% of causal genes at Recall@1 and remain below \textasciitilde55\% even at Recall@10, whereas classical tools reach \textasciitilde50--75\% and LA-MARRVEL attains \textasciitilde78\% at Recall@1 and \textasciitilde90--95\% by Recall@10, demonstrating that an LLM alone is insufficient and that coupling it to a high-recall first-stage ranker is essential for clinically useful performance. \\
    \textbf{(B)} Recall@$K$ (K=1-10) is shown for LA-MARRVEL, AI-MARRVEL, Exomiser, and LIRICAL across the BG, DDD, and UDN cohorts. LA-MARRVEL consistently achieves the highest recall at clinically salient low K (Top-1/Top-3) while maintaining near-ceiling recall by Top-10, demonstrating improved prioritization of causal genes over both the first-stage ranker and established phenotype-driven tools. \\
  }
  \label{fig:F2}
\end{figure}

Across cohorts, prompt-only LLM ranking achieved only \textasciitilde12--15\% Recall@1, versus \textasciitilde50\% for Exomiser\citep{Jacobsen2022}, \textasciitilde31\% for LIRICAL\citep{Robinson2020}, and \textasciitilde78\% for LA-MARRVEL (\Cref{fig:F2A}). We compared two prompt-only LLM settings (Claude and Claude-Thinking), two widely used phenotype-driven baselines (Exomiser and LIRICAL), and our internal first-stage baseline (AI-MARRVEL\citep{Mao2024}) to represent standalone LLM retrieval, established clinical prioritizers, and the upstream ranker used by LA-MARRVEL. The gap persists at wider review windows: at Top-5/Top-10, LLM-only recall is \textasciitilde37--55\% versus \textasciitilde75--85\% for Exomiser/LIRICAL and \textasciitilde90--95\% for LA-MARRVEL; even at Recall@100, LLM-only performance plateaus around \textasciitilde70--76\%.

In this setting, prompt-only LLM ranking showed lower retrieval performance than comparator pipelines for rare-disease gene discovery. A high-recall first stage is therefore required, and LA-MARRVEL's rank-then-rerank design combines that stage with engineered context to preserve coverage while allowing the LLM to improve prioritization at clinically relevant top ranks.

\subsection{Performance Across Three Independent Cohorts}
Across BG, DDD, and UDN, LA-MARRVEL achieved the highest Recall@$K$ at every $K=1$--$10$, with the largest gains at Top-1/Top-3 (\Cref{fig:F2B}). At the salient low-$K$ thresholds, LA-MARRVEL improved by approximately \textit{5--20 percentage points over AI-MARRVEL}, \textit{10--30 points over Exomiser}, and \textit{45--50 points over LIRICAL}. The advantage is especially pronounced on the BG dataset, where LA-MARRVEL reaches \textasciitilde80\% by Top-1 while the next-best method trails several points lower. In the DDD and UDN datasets, LA-MARRVEL approaches a ceiling rapidly, surpassing \textasciitilde95\% by Top-5 and maintaining a small but consistent margin over AI-MARRVEL at all $K$.
Performance at larger $K$ also favors LA-MARRVEL. By Top-10, it attains \textit{\textasciitilde90--95\% recall on all three cohorts}, compared with \textit{\textasciitilde75--85\% for Exomiser} and \textit{\textasciitilde70--80\% for LIRICAL}. These results indicate that LA-MARRVEL not only surfaces the correct diagnosis more often at the very top of the list --- where it matters most for clinician time and follow-up testing --- but also maintains superior coverage when a wider gene set is considered.
Together, these findings demonstrate that LA-MARRVEL shows strong performance across diverse cohorts and outperforms established gene-prioritization algorithms, offering both higher precision at the top ranks and near-ceiling recall with modest review effort.

\subsection{Predominantly Positive Net Effect Overall}

\begin{figure}
\centering
\includegraphics[width=1\textwidth]{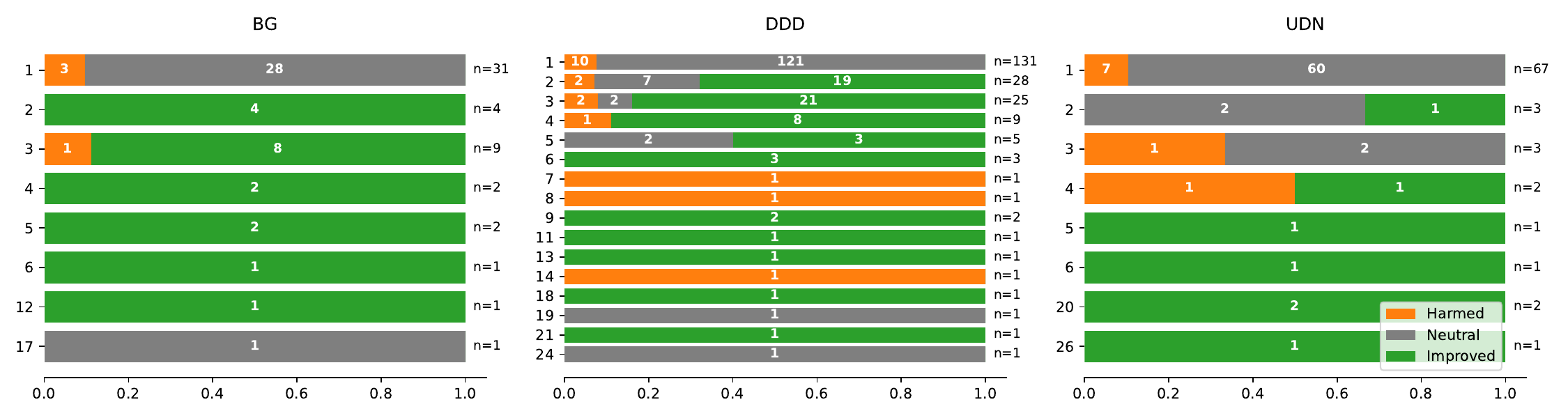}
\caption{\textcolor{Highlight}{
\textbf{Ratio of Improved and Harmed Case by Original Rank}} \\
Outcome distribution by original rank for three datasets (BG, DDD, UDN). Although \textit{Harmed} cases are observed for some ranks (orange), the number of \textit{Improved} cases (green) is consistently larger, indicating a predominantly positive effect overall; \textit{Neutral} outcomes are shown in gray. Counts are annotated within bars, and $n$ denotes the total number of cases per rank.}
\label{fig:F3}
\end{figure}

Across BG, DDD, and UDN, improved cases outnumber harmed cases in most baseline-rank bins, while rank-1 harmed fractions are BG \textasciitilde9.7\% (3/31), DDD \textasciitilde7.6\% (10/131), and UDN \textasciitilde10.4\% (7/67) (\Cref{fig:F3}). Counts within bars and the per-rank totals ($n$) indicate that some rank-level percentages are based on small samples.

Key observations are:
\begin{itemize}[noitemsep,topsep=0pt]
    \item Improvement is widespread but not universal across ranks. BG and UDN show many rank bins where improved outcomes dominate, including several deeper-rank bins that are fully improved; however, many of these bins have small $n$. DDD remains net positive overall but shows mixed distributions at multiple ranks (improved, harmed, and neutral all present in different bins).
    \item Harmed outcomes appear at some ranks, including rank 1. At original rank 1, harmed fractions are BG \textasciitilde 9.7\% (3/31), DDD \textasciitilde 7.6\% (10/131), and UDN \textasciitilde 10.4\% (7/67). Additional harmed-only or harmed-heavy bins are present at selected non-top ranks (notably in DDD and UDN), indicating that demotions can occur beyond the highest baseline ranks.
\end{itemize}

In summary, \Cref{fig:F3} supports a predominantly positive net effect overall, with improved cases generally exceeding harmed cases despite rank-specific harmed and neutral outcomes.

\subsection{Effect of Knowledge-Grounded Prompt Components}

\begin{figure}
\centering
\includegraphics[width=1\textwidth]{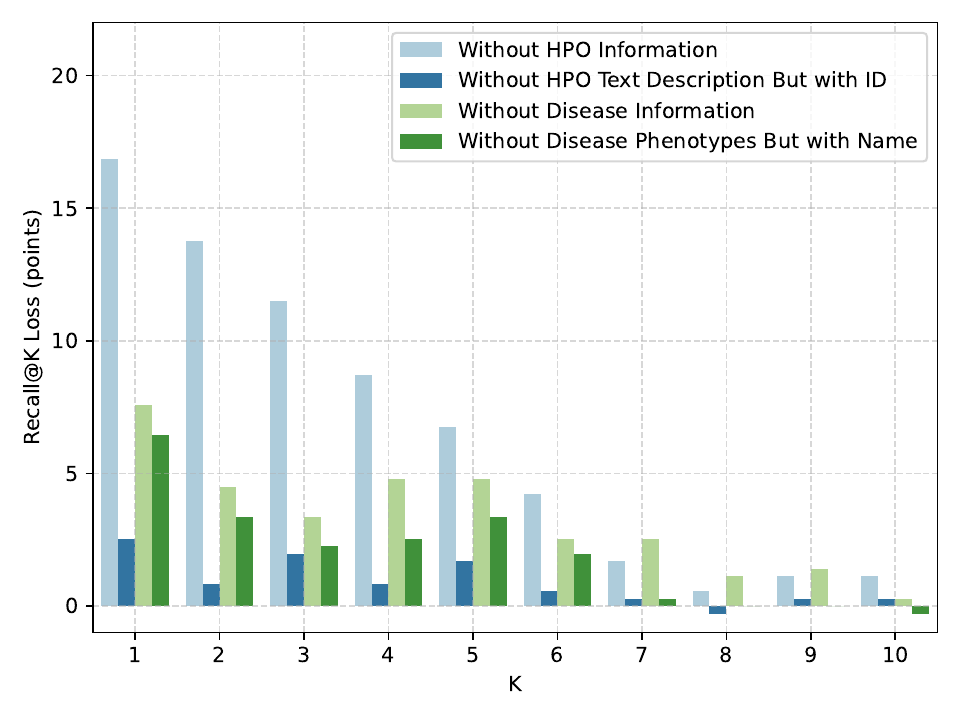}
\caption{\textcolor{Highlight}{
\textbf{Impact of Context Ablation on Recall@$K$ Loss(\%) Performance}} \\
An ablation study analyzing the impact of removing specific information context on Recall@$K$ Loss. The results show that \textbf{removing Patient Phenotype} (``Without HPO Information'') is the most harmful modification. \textbf{Removing Disease Phenotypes} is roughly as harmful as removing Total Disease Information for $K=1\text{--}6$, but shows no performance downgrade for $K=7\text{--}10$. In comparison, \textbf{removing Patient HPO Text} (``Without HPO Text Description But with ID'') is significantly less harmful.
}
\label{fig:F5}
\end{figure}

Prompt ablations show that removing patient phenotype information causes the largest Recall@$K$ loss, with disease-phenotype removal the next most detrimental condition (\Cref{fig:F5}). This indicates that patient phenotype grounding is the strongest driver in our ablation.

We further find that removing disease phenotypes is nearly as detrimental as removing whole disease information. This suggests that disease names alone often lack enough signal for model reasoning. Many disease names are not self-explanatory and are difficult for an LLM to interpret without explicit phenotype context.

For example, ``Stormorken syndrome'' as a label carries limited standalone clinical signal. In contrast, phenotype descriptors (e.g., short stature, myopathy, epistaxis, deeply set eyes, and short philtrum) provide actionable diagnostic context. Together, these findings show that knowledge-grounded prompt generation improves LA-MARRVEL by anchoring both patient and disease representations in explicit phenotype information, rather than relying on disease labels alone.

\subsection{Effect of Candidate-Set Size on Recall}

\begin{figure}
\centering
\includegraphics[width=1\textwidth]{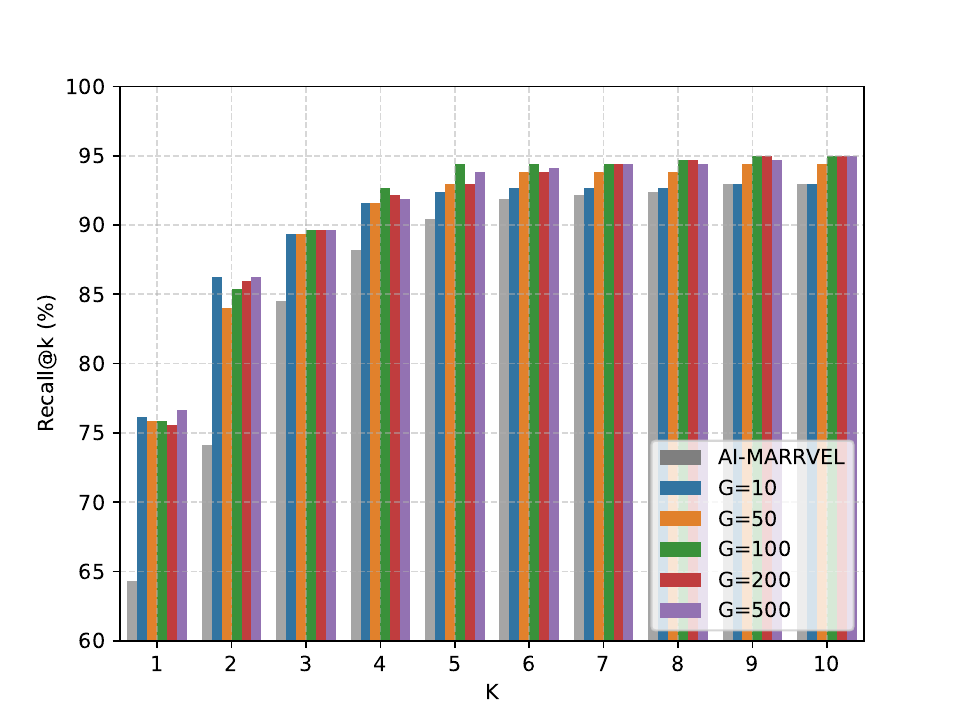}
\caption{\textcolor{Highlight}{
\textbf{Performance across different Top-$G$ candidate sets}} \\
Recall@$K$ (\%) is shown for $k \in \{1, \dots, 10\}$ across various candidate set sizes ($G$). While increasing the number of top genes from $G=10$ to $G=100$ results in a substantial boost in recall, performance begins to saturate beyond $G=100$. This indicates that the causal gene is typically captured within the top 100 candidates; further expanding the gene pool introduces diminishing returns while increasing the noise for the LLM. Based on this trade-off, $G=100$ was selected as the final operating setting. For $k \ge 7$, all $G$ configurations converge toward a recall of approximately 95\%.
}
\label{fig:F6}
\end{figure}

Increasing candidate-set size improves coverage: by $K=10$, recall rises from \textasciitilde92.8\% at $G=10$ to \textasciitilde94.4--94.9\% at $G=50$--$500$, matching or surpassing AI-MARRVEL (\textasciitilde92.8\%; \Cref{fig:F6}). The same pattern is visible from $K=5$ onward, where larger candidate pools maintain a \textasciitilde1.5--2.5 point advantage over the smaller-$G$ setting.

At the same time, top-rank concentration is less stable as candidate pools expand. Adding more genes increases competition from distractor candidates, so gains at top-$K$ coverage are generally stronger than gains at top-1 concentration.

Performance gains also saturate. Moving from $G=10$ to $G=50$ gives a clear jump (e.g., at $K=10$, \textasciitilde92.8\% to \textasciitilde94.4\%), whereas expanding from $G=100$ to $G=500$ adds only marginal benefit (\textasciitilde94.9\% to \textasciitilde94.9\%). Together, these results suggest a practical operating point around $G=50$--$100$: small enough to limit ranking noise, but large enough to retain strong diagnostic coverage. Based on this trade-off analysis, we use $G=100$ as the default final setting in subsequent experiments. We next examine a separate axis, the effect of ranked-voting aggregation versus a single run.

\subsection{Ranked Voting vs Single-Run Inference}

\begin{figure}
\centering
\includegraphics[width=1\textwidth]{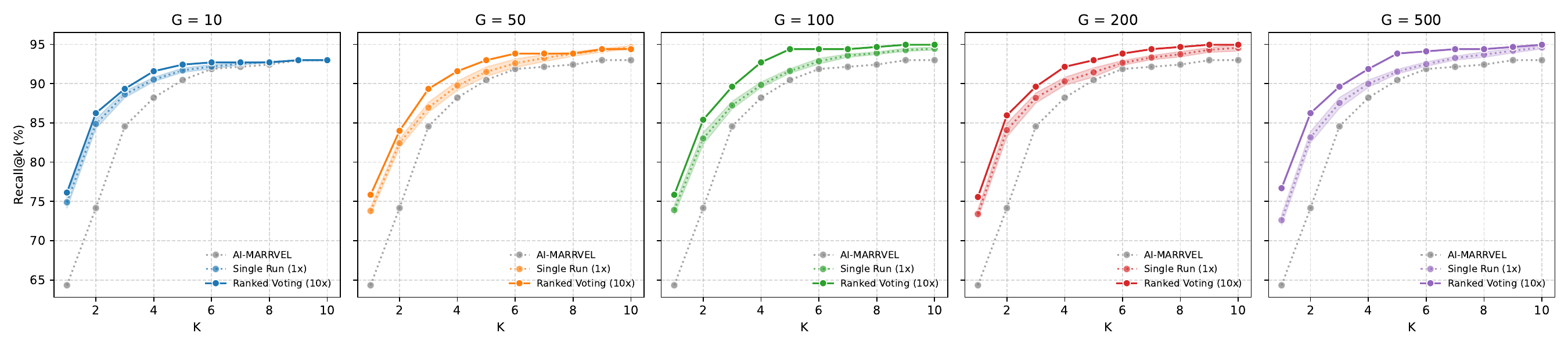}
\caption{\textcolor{Highlight}{
\textbf{Recall Curve of Performance By Differing Top G Genes and N Outputs}} \\
Scaling Analysis of Recall Performance Across Varying Gene Thresholds ($G$) and LLM Ensemble Methods. Each panel illustrates the Recall@$K$ (for $k=1$ to $10$) across increasing candidate gene pool sizes ($G = 10$ to $500$). The baseline AI-MARRVEL (dotted gray line) is compared against Single Run (1x) (dashed line) and Ranked Voting (10x) (solid line) LLM-augmented configurations. \\
Notably, the Ranked Voting (10x) method maintains superior performance across all thresholds, demonstrating its robustness even as the candidate pool grows larger and more complex.
}
\label{fig:F7}
\end{figure}

Across all tested gene-set sizes ($G \in \{10, 50, 100, 200, 500\}$) and all $k$, Ranked Voting (10x) consistently exceeds Single Run (mean of 10 independent 1x runs with 95\% confidence intervals; \Cref{fig:F7}). The Ranked Voting curves also remain above both Single Run and the baseline AI-MARRVEL curves across values of $k$, indicating that ensemble aggregation provides a robust performance floor in high-variance settings. Based on the $G$ trade-off analysis above, we use $G=100$ as the default in subsequent experiments.

The separation between Ranked Voting and Single Run remains pronounced in larger candidate sets. This pattern supports our observation that larger pools increase per-run stochasticity, while voting dampens this variability and surfaces true positives more reliably. Consequently, even when absolute recall changes only modestly with larger $G$, ensembling was associated with more stable top-rank performance.

\subsection{Explainable Traces in Representative Cases}

\begin{figure}
\centering
\includegraphics[width=1\textwidth]{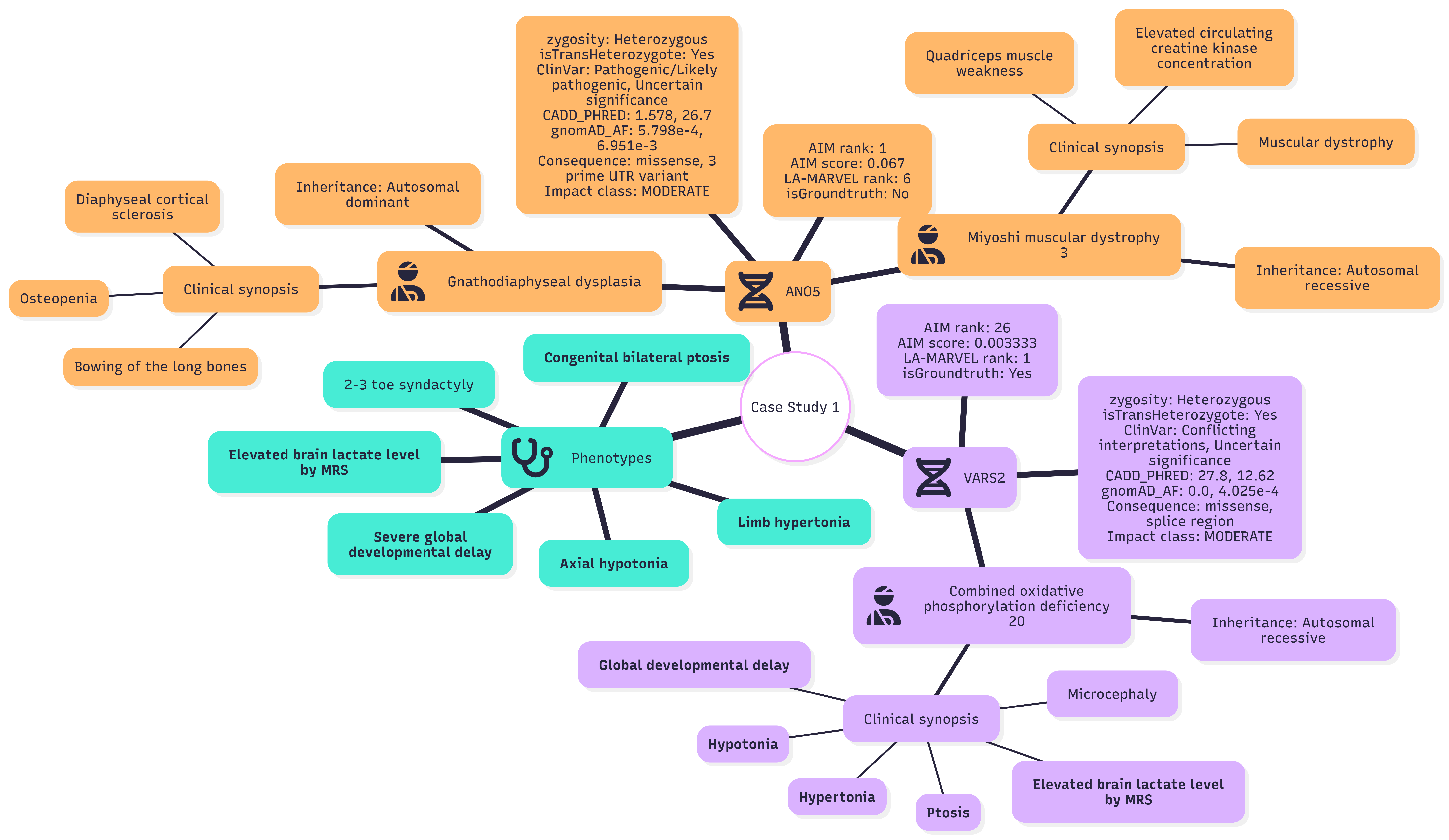}
\caption{\textcolor{Highlight}{
\textbf{The Overview of Case Study 1}} \\
Each node represents either the patient, a gene, or a phenotype group. The
\textit{phenotype} node contains the list of phenotypes observed in the
case. Gene nodes indicate candidate genes that are reprioritized (promoted) or deprioritized (demoted) by LA-MARRVEL. Nodes directly connected to gene nodes represent known relationships (e.g., gene--disease or gene--phenotype links) supporting or weakening their candidacy. Items shown in \textbf{bold} indicate matches to the patient's observed phenotypes.
}
\label{fig:F8}
\end{figure}

\begin{figure}
\centering
\includegraphics[width=1\textwidth]{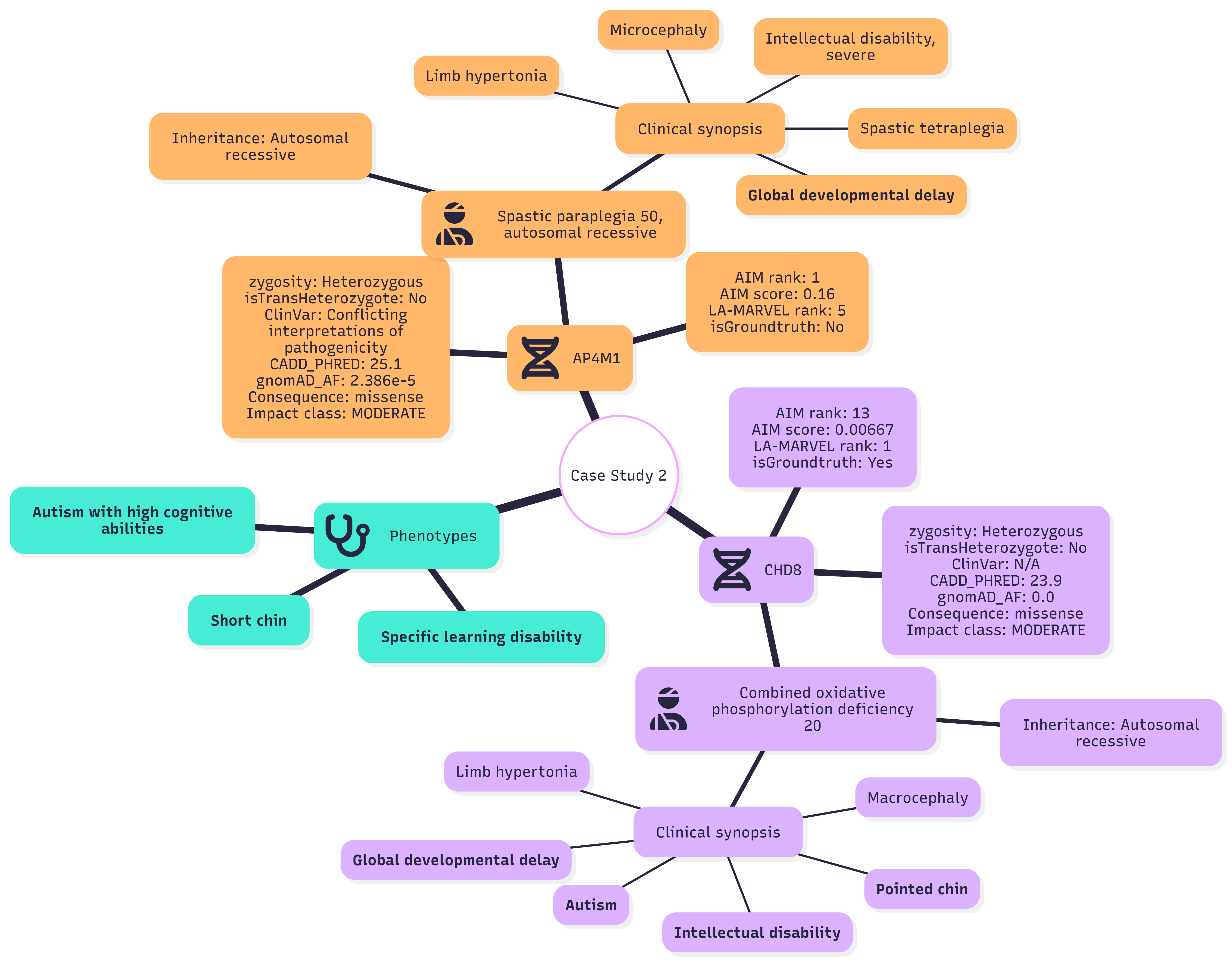}
\caption{\textcolor{Highlight}{
\textbf{The Overview of Case Study 2}} \\
Each node represents either the patient, a gene, or a phenotype group. The
\textit{phenotype} node contains the list of phenotypes observed in the
case. Gene nodes indicate candidate genes that are reprioritized (promoted) or deprioritized (demoted) by LA-MARRVEL. Nodes directly connected to gene nodes represent known relationships (e.g., gene--disease or gene--phenotype links) supporting or weakening their candidacy. Items shown in \textbf{bold} indicate matches to the patient's observed phenotypes.
}
\label{fig:F9}
\end{figure}

In representative cases, LA-MARRVEL corrected large ranking errors (VARS2: rank~26 $\rightarrow$ rank~1; CHD8: rank~13 $\rightarrow$ rank~1), and the accompanying traces document each evidence step (\Cref{fig:F8,fig:F9}). Each panel shows ACMG evidence codes --- the American College of Medical Genetics and Genomics criteria for classifying pathogenic variants\citep{Richards2015} --- zygosity checks, and how well the phenotype matches HPO terms. This trace follows the manual curation process and records the reasons for any changes in rank. Case-rationale excerpts below include model-generated wording and are shown to illustrate the system’s explanation trace.

\subsubsection{Case Study 1: VARS2 reprioritized to highest rank}

In this case, AI-MARRVEL put \textbf{ANO5}\citep{HGNC_ANO5} at rank~1 and \textbf{VARS2}\citep{HGNC_VARS2} at rank~26, even though the true diagnosis is linked to \textbf{VARS2} (\Cref{fig:F8}). LA-MARRVEL's reasoning trace walked through the same evidence a human would check and corrected the order.
The trace below is an excerpt; non-critical lines are omitted for readability.

\begin{promptbox}{LLM Explainer Trace for Case Study 1}
\begin{small}

\tracegap
\vspace{0.5\baselineskip}

\begin{normalsize} Priority 1: VARS2 - Combined oxidative phosphorylation deficiency 20 \end{normalsize}

ACMG Classification: Likely Pathogenic
\begin{itemize}[noitemsep,topsep=0pt]
  \item PM2: Both variants extremely rare (AF: 0.0, 0.0004025)
  \item PM3: Trans-heterozygous configuration in AR disorder
  \item PP3: High CADD scores (27.8, 12.62) predict damaging effects
  \item PP1: Excellent phenotype match (9/10)
\end{itemize}

Phenotype Match: Outstanding --- hypertonia, microcephaly, cerebellar hypoplasia, corpus callosum hypoplasia, elevated brain lactate by MRS, global developmental delay, hypotonia, hypertrophic cardiomyopathy, ptosis, muscle weakness, seizures, lactic acidosis. This perfectly explains the decreased PDH activity as secondary mitochondrial dysfunction.

\vspace{0.5\baselineskip}
\traceomittedgenes
\vspace{0.5\baselineskip}

\begin{normalsize} Lower Priority Genes: \end{normalsize}

\begin{itemize}[noitemsep,topsep=0pt]
  \item ANO5: Muscular dystrophy phenotype doesn't match
  \item IFIH1: Some variants benign, phenotype less consistent 
  \item BICD2: Spinal muscular atrophy, only one variant
  \item Other genes: Either benign variants, poor phenotype matches, or common variants
\end{itemize}

\tracegap
\end{small}
\end{promptbox}

The rationale text below reflects model-generated prioritization language and is presented to illustrate how the system explains ranking changes.

\paragraph{ANO5: Rationale for Demotion Despite Plausible Genotype.} Although \textbf{ANO5} contains a missense variant and a 3'UTR variant annotated as pathogenic in ClinVar, the aggregate evidence remains equivocal and the case-level phenotype fit is limited. In silico support is inconsistent (CADD 1.578 vs.\ 26.7), and while the observed allele frequencies (0.0005798 and 0.00695093) are compatible with uncommon variation, the inferred trans-heterozygous configuration most closely aligns with established \emph{ANO5}-associated muscular dystrophy/myopathy mechanisms. By contrast, this patient’s predominant manifestations are neurological and metabolic (e.g., hypertonia, microcephaly, cerebellar and corpus callosum hypoplasia, elevated brain lactate), which do not correspond to the canonical \textbf{ANO5} phenotypic spectrum. Thus, despite a potentially plausible genetic configuration, the weak concordance with the primary presenting phenotype would be expected to drive LA-MARRVEL to deprioritize \textbf{ANO5} relative to candidates with stronger phenotype matching.

\paragraph{VARS2: Rationale for Prioritization with Outstanding Phenotypic Matching.} \textbf{VARS2} represents the most compelling candidate because the molecular findings and clinical presentation converge on a coherent mitochondrial / metabolic disease mechanism. The gene contains a missense variant and a splice-region variant with discordant ClinVar assertions; however, in silico prediction suggests substantial functional impact for at least one allele (CADD 27.8). The second allele is very rare (0.0004025) and shows a moderate predicted effect (CADD 12.62), together supporting the plausibility of a bi-allelic (recessive) contribution. Critically, phenotype  concordance with \textbf{VARS2}-associated oxidative phosphorylation deficiency is strong: hypertonia, microcephaly, cerebellar and corpus callosum hypoplasia, and---most notably---elevated brain lactate on MRS are classic indicators of mitochondrial/respiratory chain dysfunction. Given this combination of a plausible recessive genotype and an ``outstanding'' match to the case's neurologic and metabolic signature, phenotype-driven prioritization (e.g., LA-MARRVEL) would be expected to elevate \textbf{VARS2} toward the top of the candidate list, potentially superseding genes supported primarily by computational prediction or less specific phenotypic overlap.

\subsubsection{Case Study 2: CHD8 reprioritized to highest rank}

In this case, AI-MARRVEL initially ranked \textbf{AP4M1}\citep{HGNC_AP4M1} first and \textbf{CHD8}\citep{HGNC_CHD8} thirteenth, even though the ground truth is linked to \textbf{CHD8} (\Cref{fig:F9}). LA-MARRVEL's explainable trace re-evaluated the molecular, inheritance, and phenotype evidence and corrected the ranking.
The trace below is an excerpt; non-critical lines are omitted for readability.

\begin{promptbox}{LLM Explainer Trace for Case Study 2}
\begin{small}

\tracegap

\begin{normalsize} 1. CHD8 - HIGHEST PRIORITY 5/5 \end{normalsize}

ACMG Classification: Likely Pathogenic
\begin{itemize}[noitemsep,topsep=0pt]
  \item Variant: Missense, CADD 23.9, gnomadAF 0.0 (absent from population)
  \item ACMG Evidence:
  \begin{itemize}[noitemsep,topsep=0pt]
    \item PM1: Missense in critical autism gene
    \item PM2: Absent in population databases 
    \item PP3: High CADD score (23.9)
    \item PS4: Phenotype highly specific for known disorder
  \end{itemize}
  \item Phenotype Match: EXCELLENT --- Disease is ``Intellectual developmental disorder with autism and macrocephaly''
  \begin{itemize}[noitemsep,topsep=0pt]
    \item Direct match: Autism (core feature)
    \item Related: Pointed chin (opposite but related facial dysmorphism to patient's short chin)
    \item Compatible: Variable intellectual impact allows for ``high cognitive abilities''
  \end{itemize}
  \item Inheritance: Autosomal dominant --- fits heterozygous state perfectly
\end{itemize}

\vspace{0.5\baselineskip}
\traceomittedgenes
\vspace{0.5\baselineskip}

\begin{normalsize} Single heterozygous variants for AR conditions (not pathogenic): \end{normalsize}

\begin{itemize}[noitemsep,topsep=0pt]
  \item AP4M1, DYNC2LI1, PLXND1, ARPC1B, ADAMTS18, LAMA3 --- All require biallelic variants
\end{itemize}

\tracegap
\end{small}
\end{promptbox}

\paragraph{AP4M1: Rationale for Demotion Due to Incomplete Recessive Evidence.} Although \textbf{AP4M1} carries a rare missense variant with strong in silico deleteriousness (CADD 25.1; gnomAD 2.386$\times$10$^{-5}$) and ClinVar reports conflicting interpretations, the overall genetic model and case-level phenotype fit are weak. \textbf{AP4M1} is primarily associated with spastic paraplegia 50, an autosomal recessive disorder, yet this case shows only a single heterozygous variant without evidence of a second allele (i.e., not compound heterozygous). In that context, a lone heterozygous missense change is insufficient to support a recessive diagnosis in the absence of additional supporting evidence (second hit, segregation, functional data). Moreover, the  phenotype concordance is limited---there is no clear match to the noted features (e.g., autism/learning disability/short chin) under the provided summary. Therefore, despite rarity and a high CADD score, the discordant inheritance model and weak phenotype match would be expected to lead LA-MARRVEL to rank AP4M1 below candidates demonstrating both compatible inheritance and stronger clinical concordance.

\paragraph{CHD8: Rationale for Prioritization Under a Plausible Dominant Model.} \textbf{CHD8} is a strong candidate because its variant profile and inheritance pattern align well with a gene where single-allele disruption is known to cause disease, and the case summary suggests clinically relevant overlap. The variant is heterozygous, ultra-rare in gnomAD (0.0), and supported by high in silico deleteriousness (CADD 23.9), collectively increasing plausibility for pathogenicity despite the absence of a ClinVar entry. Importantly, \textbf{CHD8} is associated with an autosomal dominant neurodevelopmental disorder characterized by intellectual disability and autism-related features, which corresponds to the reported developmental delay and intellectual disability in this case. Compared with recessive candidates represented by only one detected allele, \textbf{CHD8} is consistant with a dominant model in which a single heterozygous (likely damaging) variant can explain the core neurodevelopmental manifestations. Consequently, phenotype- and inheritance-aware prioritization (e.g., LA-MARRVEL) would be expected to elevate \textbf{CHD8} relative to genes with weaker phenotype concordance or an unsupported recessive genotype configuration.

\section{Discussion}

In this study, we demonstrate that a language-grounded reranking framework can improve the robustness and interpretability of rare-disease gene prioritization. This study suggests a practical design choice for rare disease diagnosis: LLMs are most effective as second-stage rerankers on top of high-recall retrieval, rather than as standalone diagnostic tools. Across three independent cohorts (BG, DDD, and UDN), LA-MARRVEL improved top-$K$ performance while preserving first-stage detection coverage. Prompt-only LLM ranking showed lower causal-gene recovery rate (about 12--15\% Recall@1 and 37--55\% at Top-5/Top-10), whereas established baselines (Exomiser and LIRICAL) maintained higher recall. LA-MARRVEL then achieved the best performance across $K=1$--$10$, with larger gains at Top-1/Top-3 and near-ceiling recall by Top-10. These findings align with prior observations that existing prioritization tools achieve high recall but leave substantial ranking uncertainty for clinical interpretation. For example, in 4,877 known diagnoses from the 100,000 Genomes Project, Exomiser achieved a recall of 0.92 and a precision of 0.18 when reviewing the top five candidates \citep{Jacobsen2022}. In an independent cohort, AI-MARRVEL reported a precision--recall AUC of 0.8248; at high recall (\(\sim 0.9\)), precision was \(\sim 0.3\) \citep{Mao2024}. Prompt-only LLM ranking is sensitive to input ordering and literature prominence \citep{Neeley2024}, rather than systematically weighting phenotypic concordance or ACMG evidence criteria. LA-MARRVEL addresses this gap through a modular pipeline that preserves recall while improving precision among top-ranked candidates. In practice, this reduces the candidate set to a clinically manageable subset, where small changes in rank order can meaningfully influence downstream review and decision-making (e.g., VARS2: rank 26th to 1st; CHD8: rank 13th to 1st).

Two factors likely drive these gains: (i) evidence-rich context construction and (ii) deterministic aggregation of multiple LLM-generated rank lists. Rare disease evidence is fragmented across many annotation layers; LA-MARRVEL's prompt composer integrates these signals into compact, case-specific context, enabling direct variant-bearing gene comparison. Ablation results support this mechanism: removing patient phenotypes caused the largest Recall@$K$ drop, and removing disease-phenotype relationship information was nearly as harmful as removing disease information entirely. Disease names alone often provide limited inferential signal, especially for eponymous or non-compositional labels, whereas phenotype descriptors offer more clinically actionable semantics. In this setting, the composer-powered prompt functions as structured case representation, not instruction alone. A second factor is stability through aggregation. Single-run LLM outputs can vary, especially when evidence is ambiguous or candidate pools are large. LA-MARRVEL reduces this variability and improves accuracy through repeated inference \citep{WangSelfCons2023} and deterministic Ranked Pairs (Tideman) aggregation \citep{Tideman1987}. Empirically, Ranked Voting (10$\times$) outperformed single-run inference across tested gene-set sizes ($G \in {10, 50, 100, 200, 500}$) and across $k$, yielding a more stable performance floor. This is important for clinical translation, where average gains are insufficient if run-to-run behavior is unstable. Deterministic aggregation is therefore not only a performance optimization but also a reliability mechanism.

However, ranking robustness alone does not determine practical performance; the scope of the candidate set produced by first-stage retrieval also shapes how reranking affects outcomes. Candidate-set analysis further clarifies trade-offs between coverage and concentration. Increasing candidate-set size $G$ improved coverage (e.g., by $K=10$, recall rose from \textasciitilde92.8\% at $G=10$ to \textasciitilde94.4--94.9\% at $G=50$--$500$), but larger pools add distractors, so gains were generally stronger for Top-5/Top-10 than for Top-1. Performance also saturated: improvement from $G=10$ to $G=50$ was substantial, whereas expansion from $G=100$ to $G=500$ yielded limited additional benefit. These patterns support an operating range of $G=50$--$100$ (consistent with the default $G=100$), and indicate that candidate expansion is a tunable parameter with diminishing returns.


Moreover, LA-MARRVEL emphasizes transparency and explainability. In representative cases, reranking corrected substantial ordering errors (e.g., VARS2: rank 26th $\rightarrow$ 1st; CHD8: rank 13th $\rightarrow$ 1st) and generated auditable reasoning traces resembling key manual review steps, including phenotype matching, inheritance and zygosity assessment, and ACMG-like evidence descriptors \citep{Richards2015}. These traces do not guarantee correctness but provide structured artifacts that allow users to evaluate why one candidate is prioritized over another and to identify mechanistically inconsistent genes. Such traceability facilitates critical review and documentation within existing clinical interpretation workflows.


Existing gene-prioritization frameworks such as AI-MARRVEL, Exomiser, and LIRICAL rely on statistical or conventional machine-learning models to rank variant-bearing genes and achieve strong retrieval performance \citep{Mao2024,Jacobsen2022,Robinson2020}. However, these approaches primarily operate on structured features and do not integrate heterogeneous case-specific evidence at the level of language-level reasoning. In parallel, literature-driven retrieval methods using NLP accelerate candidate generation \citep{Birgmeier2020}, but rely on predefined feature engineering that constrains the representation of phenotypic and genomic evidence, and still require manual reconciliation of competing hypotheses \citep{Mao2024}. Recent agentic or tool-augmented systems couple retrieval with LLM orchestration \citep{Zhao2026DeepRare}, enabling broader reasoning but increasing complexity as candidate pools expand. LA-MARRVEL instead adopts a reranking paradigm: high-recall retrieval is preserved upstream, while language-based synthesis is applied only to a constrained candidate set. This design concentrates computation on comparative interpretation rather than search and remains compatible with existing variant interpretation workflows.


LA-MARRVEL improved recall, stability, and robustness overall, but gains were not uniform across cases. Net-effect analysis showed more improvements than degradations across most baseline-rank bins; however, adverse rank shifts occurred even when the causal gene was initially ranked first (BG \textasciitilde9.7\%, DDD \textasciitilde7.6\%, UDN \textasciitilde10.4\%). This is clinically relevant because aggregate gains can obscure case-level demotions. Therefore, LA-MARRVEL is best interpreted as decision support rather than automated adjudication, and rankings should remain subject to clinical review.

Several limitations constrain interpretation. First, evaluation was retrospective on diagnosed cohorts; performance may differ in prospective workflows where phenotypes evolve and documentation is incomplete. Second, performance depends on input quality, including completeness of HPO annotations and variant metadata, and on correct disease-model assumptions. Third, transparency and explainability do not ensure factual correctness: medical LLMs remain susceptible hallucination and citation-faithfulness errors \citep{Asgari2025,Wu2025}, so systems should communicate uncertainty and scope explicitly. Finally, repeated inference and ranked voting increase computational cost and latency, which may limit use in resource-constrained settings or rapid-turnaround settings.

These limitations motivate two complementary directions for future work. The first area of focus is the design of models and underlying algorithms. Prompt construction could incorporate additional structured evidence, including richer gene/variant annotations, phenotype hierarchy information, family history and segregation evidence, and selected content from unstructured notes. Future work should prioritize context needed for ACMG/AMP-based variant interpretation, including ClinGen SVI recommendations \citep{ClinGenSVIGuidance2026} and VCEP gene/disease-specific specifications (e.g., hearing loss and MYH7 cardiomyopathy) \citep{Oza2018HLVCEP,Kelly2018MYH7VCEP}, within a single-pass prompting framework to avoid extra orchestration overhead. A related direction is adaptive ensembling. Ranked voting appears increasingly useful as candidate sets grow but is costly when applied uniformly. A natural extension is to adapt the number of runs $N$ by case-level uncertainty, using adaptive stopping when responses converge \citep{LeeConSol2025}. Separately, candidate-set size $G$ can be adjusted, with smaller settings for stable cases and larger settings when evidence is weak or inconsistent. This conditional computation could preserve recovery in difficult cases while reducing cost in straightforward ones.


The second direction concerns clinical deployment, human–system interaction, and impact evaluation. For safe clinician-in-the-loop integration, explanation traces should evolve toward claim-level, source-grounded artifacts, with explicit links between each assertion and its supporting evidence, clear uncertainty and conflict indicators, and compatibility with institution-specific reporting formats. Transparency alone is insufficient: prior work shows that explainability must be paired with user onboarding \citep{Chaing2021,Passi2022OverrelianceAI}, explicit communication of system scope and limitations \citep{Yin2019,Passi2022OverrelianceAI}, and workflow safeguards designed to mitigate automation bias \citep{Goddard2012AutomationBias}. Beyond interface design, rigorous impact evaluation is essential. Assessment should extend beyond retrospective ranking metrics to prospective reader studies that compare interpretation accuracy, review time, and resource utilization with and without LA-MARRVEL support across clinician roles and levels of experience \citep{Novak2024,Lyell2025}. Such studies would clarify not only performance effects but also how language-grounded reranking shapes decision behavior in real-world settings.

In summary, LA-MARRVEL demonstrated that language-based reranking can improve rare disease candidate prioritization when embedded in a high-recall pipeline and stabilized by deterministic aggregation. Benefits are concentrated at the top of the ranked list, where prioritization decisions are most consequential, while still providing inspectable rationale for adjudication. As diagnostic systems increasingly integrate annotation, phenotype normalization, retrieval, and LLM reasoning, modular approaches that preserve first-stage recall and improve top-of-list precision may remain practical. Future work should prioritize stronger provenance, adaptive compute and candidate control, and prospective human-centered evaluation to determine how measured gains translate into reliable, safe clinical impact.

\section{Materials and Methods}

\subsection{Datasets}
We used three clinical cohorts from AI-MARRVEL to ensure a consistent, apples-to-apples comparison with prior work. Summary statistics are shown in \Cref{tab:1}. In brief:

\begin{itemize}[noitemsep,topsep=0pt]
\item \textbf{BG (Baylor Genetics).} 63 cases; on average 1469.2 variant-bearing genes, 10.59 HPO terms, and 1.175 causal genes per case, spanning 74 unique causal genes. Curated in-house; accessibility is \emph{Internal}.
\item \textbf{DDD (Deciphering Developmental Disorders).} 214 cases; on average 811.5 variant-bearing genes, 7.38 HPO terms, and 1.000 causal genes per case, spanning 116 unique causal genes. Curated by AMELIE\citep{Birgmeier2020}; accessibility is \emph{Restricted}.
\item \textbf{UDN (Undiagnosed Diseases Network).} 90 cases; on average 1072.2 variant-bearing genes, 43.98 HPO terms, and 1.044 causal genes per case, spanning 93 unique causal genes. Curated in-house; accessibility is \emph{Restricted}.
\end{itemize}

We adopt the original curation provided by each source; however, we excluded multi-gene cases (7 from BG and 4 from UDN), as they precluded detailed downstream investigation.

\subsection{Experiment Details}
\textbf{Model}
For all LA-MARRVEL experiments, we used Claude-4-Sonnet with Extended Thinking enabled
(\texttt{\seqsplit{anthropic.claude-sonnet-4-20250514-v1:0}}). The model was configured with a 65{,}536-token
context window and a 50{,}000-token thinking budget for each call.

\textbf{Environment} All experiments were conducted in an AWS environment configured for HIPAA workloads, using only HIPAA-eligible services for ePHI processing (including Amazon Bedrock)\citep{AWS_HIPAA_ELIGIBLE_SERVICES}. To enforce a zero-data-retention inference path for PHI, we used Bedrock's default behavior in which prompts and completions are not stored or logged and are not used to train AWS foundation models\citep{AWS_BEDROCK_DATAPROTECTION}, and we kept optional model invocation logging disabled throughout evaluation\citep{AWS_BEDROCK_INVOCATIONLOGGING}. During evaluation, the model was run in a strictly offline configuration with respect to external knowledge sources: no web search, browsing, or retrieval from external APIs was enabled or used in any of the LA-MARRVEL experiments.

\textbf{Hyperparameters} Unless otherwise specified, all decoding and sampling hyperparameters (e.g., temperature, top\_k, top\_p, and related settings) were kept at their default values provided by the Bedrock-hosted Claude-4-Sonnet endpoint\citep{AWS_CLAUDEPARAMS}. No additional tuning of these hyperparameters was performed across experiments.

\textbf{Comparisons} We used the latest available software versions at the time of experimentation. Specifically, we used Exomiser 14.1.0 (released December 10, 2024) with data version 2410; LIRICAL 2.2.0 (released June 17, 2025) with data version 2406; and AI-MARRVEL 1.0.0 (released August 29, 2024) with data version 2.0, which matches the version previously reported \citep{Mao2024}.

\subsection{First-Stage Ranks and Gene Annotations by AI-MARRVEL} 

LLMs are evolving, but at the current stage they require concise, informative context provided via a prompt. On average, a proband has variants in 800--1,500 genes, including benign ones; including all genes in a prompt is not feasible. To reduce the number of candidate genes to a feasible size, we chose the reputable tool AI-MARRVEL, a strong first-stage ranker previously validated on the BG, DDD, and UDN cohorts \citep{Mao2024}.
Unless otherwise noted, we set the candidate-set size to $G=100$ for all final LA-MARRVEL runs.

\subsection{Heterogeneous Data Inputs}

LA-MARRVEL integrates three heterogeneous input groups before reranking: (i) patient annotation (phenotypes), (ii) gene-disease annotation (HPO phenotypes in DAG form, mode of inheritance), and (iii) gene-variant annotation (variant consequence ontology, population allele frequency, ClinVar records, and CADD as computational predictors). The key bottleneck is combining these heterogeneous formats into one coherent ranking signal without dropping clinically relevant context.

\subsection{Knowledge-Grounded Prompt Composer}

AI-MARRVEL not only ranks candidate genes, it also generates variant- and gene-level annotations using Ensembl's Variant Effect Predictor (VEP) and the MARRVEL database. LA-MARRVEL then uses these to create concise gene summaries (See the below prompt template, \Cref{tab:2}, and \Cref{tab:3}), which can help preliminarily assess several ACMG/AMP criteria \citep{Richards2015}, including: variant consequence (PVS1, when predicted loss-of-function is a known disease mechanism); computational predictions such as CADD (PP3/BP4); gnomAD allele frequency (PM2/BA1, absent or rare / too common); gene-disease information (e.g., OMIM) combined with a highly specific patient phenotype (PP4); ClinVar (PP5/BP6, reputable database).

\begin{promptbox}{LA-MARRVEL Main Prompt Template}

\begin{small}
    

Given HPO and Genes with Variants, Evaluate Gene 1 by 1 based on ACMG and phenotype matching to Prioritize Pathogenic Genes.

Notes:
\begin{itemize}[noitemsep,topsep=0pt]
    \item zyg=het \& isTransHeterozygote=yes: a second pathogenic variant is present in both alleles of a gene, which meets the recessive disease criteria.
    \item zyg=hom may mean hemizygous for chromosome X if the proband is male
    \item The data is generated from whole exome sequencing; finding an additional pathogenic variant is unlikely.
\end{itemize}

\vspace{1\baselineskip}
HPO:

\textbf{\textless HPO table\textgreater}

\vspace{1\baselineskip}
Genes:

\textbf{\textless Gene summary table\textgreater}

\end{small}

\end{promptbox}

\subsection{Aggregating Partial Gene Rankings with Tideman's Method}

Ensembling LLMs at inference time \textemdash via \textit{self-consistency }(sample diverse chains of thought and pick the modal answer), structured search like Tree-of-Thoughts, or Best-of-$N$ selection with a verifier \textemdash often boosts accuracy on math, code, and reasoning benchmarks \citep{WangSelfCons2023, YaoTreeThoughts2023,WangMathShepherd2024}. In addition, sampling-efficient variants such as \textit{ConSol} use Sequential Probability Ratio Testing (SPRT) to adaptively stop once responses converge, cutting the number of samples required relative to vanilla self-consistency while preserving accuracy \citep{LeeConSol2025}.

For each LLM call, we use a two-step inference procedure: (i) generate an unconstrained free-text response (reasoning plus candidate prioritization), and (ii) deterministically extract an ordered list of gene symbols from that text to form a machine-readable partial ranking. We then complete each partial ranking with the remaining genes from \textit{InitialList} (in original order) to construct ballots for consensus aggregation. We aggregate these ballots using \textit{Ranked Pairs (Tideman)}~\citep{Tideman1987}: (1) build pairwise win counts $\text{Wins}[a,b]$ by treating any ranked item as preferred over any candidate genes; (2) for each unordered pair, form a directed victory with margin $|\text{Wins}[a,b]-\text{Wins}[b,a]|$; (3) sort directed pairs by decreasing margin, then by the winner's votes, with deterministic lexicographic tie-breaks; (4) greedily lock edges that do not create cycles; (5) output a topological order, breaking ties by Borda score (computed only from ranked positions) then by name (see \Cref{alg:rankedpairs} and \ref{alg:reorder}). Ranked Pairs is Condorcet-compliant and clone-independent, leveraging full list structure while remaining robust to near-duplicate options.


\end{doublespacing}

\section{Acknowledgments}
This work was supported by the Cancer Prevention and Research Institute of Texas (CPRIT, RP240131), the Chan Zuckerberg Initiative (2023-332162), the National Institutes of Health (NIH, U54NS093793), the Eunice Kennedy Shriver National Institute of Child Health and Human Development of the NIH (P50HD103555), the Chao Endowment, the Huffington Foundation, and the Jan and Dan Duncan Neurological Research Institute at Texas Children's Hospital.

\section{Author Competing Interests}
The authors declare no competing interests.

\newpage
\renewcommand\refname{References}
\begin{footnotesize}
\bibliographystyle{unsrtnat} 
\textnormal{\bibliography{refs/localbibliography.bib}}

@article{Neeley2024,
  title = {Survey and improvement strategies for gene prioritization with large language models},
  volume = {5},
  ISSN = {2635-0041},
  url = {http://dx.doi.org/10.1093/bioadv/vbaf148},
  DOI = {10.1093/bioadv/vbaf148},
  number = {1},
  journal = {Bioinformatics Advances},
  publisher = {Oxford University Press (OUP)},
  author = {Neeley,  Matthew B and Qi,  Guantong and Wang,  Guanchu and Tang,  Ruixiang and Mao,  Dongxue and Liu,  Chaozhong and Pasupuleti,  Sasidhar and Yuan,  Bo and Xia,  Fan and Liu,  Pengfei and Liu,  Zhandong and Hu,  Xia},
  editor = {Zhu,  Shanfeng},
  year = {2024},
  month = dec 
}

@article{Khler2020,
  title = {The Human Phenotype Ontology in 2021},
  volume = {49},
  ISSN = {1362-4962},
  url = {http://dx.doi.org/10.1093/nar/gkaa1043},
  DOI = {10.1093/nar/gkaa1043},
  number = {D1},
  journal = {Nucleic Acids Research},
  publisher = {Oxford University Press (OUP)},
  author = {K\"{o}hler,  Sebastian and Gargano,  Michael and Matentzoglu,  Nicolas and Carmody,  Leigh C and Lewis-Smith,  David and Vasilevsky,  Nicole A and Danis,  Daniel and Balagura,  Ganna and Baynam,  Gareth and Brower,  Amy M and Callahan,  Tiffany J and Chute,  Christopher G and Est,  Johanna L and Galer,  Peter D and Ganesan,  Shiva and Griese,  Matthias and Haimel,  Matthias and Pazmandi,  Julia and Hanauer,  Marc and Harris,  Nomi L and Hartnett,  Michael J and Hastreiter,  Maximilian and Hauck,  Fabian and He,  Yongqun and Jeske,  Tim and Kearney,  Hugh and Kindle,  Gerhard and Klein,  Christoph and Knoflach,  Katrin and Krause,  Roland and Lagorce,  David and McMurry,  Julie A and Miller,  Jillian A and Munoz-Torres,  Monica C and Peters,  Rebecca L and Rapp,  Christina K and Rath,  Ana M and Rind,  Shahmir A and Rosenberg,  Avi Z and Segal,  Michael M and Seidel,  Markus G and Smedley,  Damian and Talmy,  Tomer and Thomas,  Yarlalu and Wiafe,  Samuel A and Xian,  Julie and Y\"{u}ksel,  Zafer and Helbig,  Ingo and Mungall,  Christopher J and Haendel,  Melissa A and Robinson,  Peter N},
  year = {2020},
  month = dec,
  pages = {D1207–D1217}
}

@article{Murdoch2013,
  title = {The Inevitable Application of Big Data to Health Care},
  volume = {309},
  ISSN = {0098-7484},
  url = {http://dx.doi.org/10.1001/jama.2013.393},
  DOI = {10.1001/jama.2013.393},
  number = {13},
  journal = {JAMA},
  publisher = {American Medical Association (AMA)},
  author = {Murdoch,  Travis B. and Detsky,  Allan S.},
  year = {2013},
  month = apr,
  pages = {1351}
}

@article{Faye2024,
  title = {Time to diagnosis and determinants of diagnostic delays of people living with a rare disease: results of a Rare Barometer retrospective patient survey},
  volume = {32},
  ISSN = {1476-5438},
  url = {http://dx.doi.org/10.1038/s41431-024-01604-z},
  DOI = {10.1038/s41431-024-01604-z},
  number = {9},
  journal = {European Journal of Human Genetics},
  publisher = {Springer Science and Business Media LLC},
  author = {Faye,  Fatoumata and Crocione,  Claudia and Anido de Peña,  Roberta and Bellagambi,  Simona and Escati Peñaloza,  Luciana and Hunter,  Amy and Jensen,  Lene and Oosterwijk,  Cor and Schoeters,  Eva and de Vicente,  Daniel and Faivre,  Laurence and Wilbur,  Michael and Le Cam,  Yann and Dubief,  Jessie},
  year = {2024},
  month = may,
  pages = {1116–1126}
}

@article{Wise2019,
  title = {Genomic Medicine for Undiagnosed Diseases},
  journal = {The Lancet},
  author = {Wise, Amy L. and Manolio, Teri A. and Mensah, George A. and others},
  year = {2019}
}

@article{BenitoLozano2022,
  title = {Diagnostic Process in Rare Diseases: Determinants Associated With Diagnostic Delay},
  journal = {International Journal of Environmental Research and Public Health},
  author = {Benito-Lozano, Javier and Arias-Merino, Gerardo and G{\'o}mez-Mart{\'\i}nez, Mar{\'\i}a and others},
  year = {2022}
}

@article{NguengangWakap2020,
  title = {Estimating Cumulative Point Prevalence of Rare Diseases: Analysis of the Orphanet Database},
  volume = {28},
  ISSN = {1476-5438},
  url = {http://dx.doi.org/10.1038/s41431-019-0508-0},
  DOI = {10.1038/s41431-019-0508-0},
  number = {2},
  journal = {European Journal of Human Genetics},
  publisher = {Springer Science and Business Media LLC},
  author = {Nguengang Wakap, Solenne and Lambert, David M. and Olry, Anne and Rodwell, Charlotte and Gueydan, Chlo{\'e} and Lanneau, Val{\'e}rie and Murphy, Daniel and Le Cam, Yann and Rath, Ana},
  year = {2020},
  month = feb,
  pages = {165--173}
}

@article{Birgmeier2020,
  title = {AMELIE speeds Mendelian diagnosis by matching patient phenotype and genotype to primary literature},
  volume = {12},
  ISSN = {1946-6242},
  url = {http://dx.doi.org/10.1126/scitranslmed.aau9113},
  DOI = {10.1126/scitranslmed.aau9113},
  number = {544},
  journal = {Science Translational Medicine},
  publisher = {American Association for the Advancement of Science (AAAS)},
  author = {Birgmeier,  Johannes and Haeussler,  Maximilian and Deisseroth,  Cole A. and Steinberg,  Ethan H. and Jagadeesh,  Karthik A. and Ratner,  Alexander J. and Guturu,  Harendra and Wenger,  Aaron M. and Diekhans,  Mark E. and Stenson,  Peter D. and Cooper,  David N. and Ré,  Christopher and Beggs,  Alan H. and Bernstein,  Jonathan A. and Bejerano,  Gill},
  year = {2020},
  month = may 
}

@article{Garla2012,
  title = {Ontology-guided feature engineering for clinical text classification},
  volume = {45},
  ISSN = {1532-0464},
  url = {http://dx.doi.org/10.1016/j.jbi.2012.04.010},
  DOI = {10.1016/j.jbi.2012.04.010},
  number = {5},
  journal = {Journal of Biomedical Informatics},
  publisher = {Elsevier BV},
  author = {Garla,  Vijay N. and Brandt,  Cynthia},
  year = {2012},
  month = oct,
  pages = {992–998}
}

@article{Brownstein2014,
  title = {An international effort towards developing standards for best practices in analysis,  interpretation and reporting of clinical genome sequencing results in the CLARITY Challenge},
  volume = {15},
  ISSN = {1474-760X},
  url = {http://dx.doi.org/10.1186/gb-2014-15-3-r53},
  DOI = {10.1186/gb-2014-15-3-r53},
  number = {3},
  journal = {Genome Biology},
  publisher = {Springer Science and Business Media LLC},
  author = {Brownstein,  Catherine A and Beggs,  Alan H and Homer,  Nils and Merriman,  Barry and Yu,  Timothy W and Flannery,  Katherine C and DeChene,  Elizabeth T and Towne,  Meghan C and Savage,  Sarah K and Price,  Emily N and Holm,  Ingrid A and Luquette,  Lovelace J and Lyon,  Elaine and Majzoub,  Joseph and Neupert,  Peter and McCallie Jr,  David and Szolovits,  Peter and Willard,  Huntington F and Mendelsohn,  Nancy J and Temme,  Renee and Finkel,  Richard S and Yum,  Sabrina W and Medne,  Livija and Sunyaev,  Shamil R and Adzhubey,  Ivan and Cassa,  Christopher A and de Bakker,  Paul IW and Duzkale,  Hatice and Dworzyński,  Piotr and Fairbrother,  William and Francioli,  Laurent and Funke,  Birgit H and Giovanni,  Monica A and Handsaker,  Robert E and Lage,  Kasper and Lebo,  Matthew S and Lek,  Monkol and Leshchiner,  Ignaty and MacArthur,  Daniel G and McLaughlin,  Heather M and Murray,  Michael F and Pers,  Tune H and Polak,  Paz P and Raychaudhuri,  Soumya and Rehm,  Heidi L and Soemedi,  Rachel and Stitziel,  Nathan O and Vestecka,  Sara and Supper,  Jochen and Gugenmus,  Claudia and Klocke,  Bernward and Hahn,  Alexander and Schubach,  Max and Menzel,  Mortiz and Biskup,  Saskia and Freisinger,  Peter and Deng,  Mario and Braun,  Martin and Perner,  Sven and Smith,  Richard JH and Andorf,  Janeen L and Huang,  Jian and Ryckman,  Kelli and Sheffield,  Val C and Stone,  Edwin M and Bair,  Thomas and Black-Ziegelbein,  E Ann and Braun,  Terry A and Darbro,  Benjamin and DeLuca,  Adam P and Kolbe,  Diana L and Scheetz,  Todd E and Shearer,  Aiden E and Sompallae,  Rama and Wang,  Kai and Bassuk,  Alexander G and Edens,  Erik and Mathews,  Katherine and Moore,  Steven A and Shchelochkov,  Oleg A and Trapane,  Pamela and Bossler,  Aaron and Campbell,  Colleen A and Heusel,  Jonathan W and Kwitek,  Anne and Maga,  Tara and Panzer,  Karin and Wassink,  Thomas and Van Daele,  Douglas and Azaiez,  Hela and Booth,  Kevin and Meyer,  Nic and Segal,  Michael M and Williams,  Marc S and Tromp,  Gerard and White,  Peter and Corsmeier,  Donald and Fitzgerald-Butt,  Sara and Herman,  Gail and Lamb-Thrush,  Devon and McBride,  Kim L and Newsom,  David and Pierson,  Christopher R and Rakowsky,  Alexander T and Maver,  Aleš and Lovrečić,  Luca and Palandačić,  Anja and Peterlin,  Borut and Torkamani,  Ali and Wedell,  Anna and Huss,  Mikael and Alexeyenko,  Andrey and Lindvall,  Jessica M and Magnusson,  Måns and Nilsson,  Daniel and Stranneheim,  Henrik and Taylan,  Fulya and Gilissen,  Christian and Hoischen,  Alexander and van Bon,  Bregje and Yntema,  Helger and Nelen,  Marcel and Zhang,  Weidong and Sager,  Jason and Zhang,  Lu and Blair,  Kathryn and Kural,  Deniz and Cariaso,  Michael and Lennon,  Greg G and Javed,  Asif and Agrawal,  Saloni and Ng,  Pauline C and Sandhu,  Komal S and Krishna,  Shuba and Veeramachaneni,  Vamsi and Isakov,  Ofer and Halperin,  Eran and Friedman,  Eitan and Shomron,  Noam and Glusman,  Gustavo and Roach,  Jared C and Caballero,  Juan and Cox,  Hannah C and Mauldin,  Denise and Ament,  Seth A and Rowen,  Lee and Richards,  Daniel R and Lucas,  F Anthony San and Gonzalez-Garay,  Manuel L and Caskey,  C Thomas and Bai,  Yu and Huang,  Ying and Fang,  Fang and Zhang,  Yan and Wang,  Zhengyuan and Barrera,  Jorge and Garcia-Lobo,  Juan M and González-Lamuño,  Domingo and Llorca,  Javier and Rodriguez,  Maria C and Varela,  Ignacio and Reese,  Martin G and De La Vega,  Francisco M and Kiruluta,  Edward and Cargill,  Michele and Hart,  Reece K and Sorenson,  Jon M and Lyon,  Gholson J and Stevenson,  David A and Bray,  Bruce E and Moore,  Barry M and Eilbeck,  Karen and Yandell,  Mark and Zhao,  Hongyu and Hou,  Lin and Chen,  Xiaowei and Yan,  Xiting and Chen,  Mengjie and Li,  Cong and Yang,  Can and Gunel,  Murat and Li,  Peining and Kong,  Yong and Alexander,  Austin C and Albertyn,  Zayed I and Boycott,  Kym M and Bulman,  Dennis E and Gordon,  Paul MK and Innes,  A Micheil and Knoppers,  Bartha M and Majewski,  Jacek and Marshall,  Christian R and Parboosingh,  Jillian S and Sawyer,  Sarah L and Samuels,  Mark E and Schwartzentruber,  Jeremy and Kohane,  Isaac S and Margulies,  David M},
  year = {2014},
  month = mar 
}

@article{Dewey2014,
  title = {Clinical Interpretation and Implications of Whole-Genome Sequencing},
  volume = {311},
  ISSN = {0098-7484},
  url = {http://dx.doi.org/10.1001/jama.2014.1717},
  DOI = {10.1001/jama.2014.1717},
  number = {10},
  journal = {JAMA},
  publisher = {American Medical Association (AMA)},
  author = {Dewey,  Frederick E. and Grove,  Megan E. and Pan,  Cuiping and Goldstein,  Benjamin A. and Bernstein,  Jonathan A. and Chaib,  Hassan and Merker,  Jason D. and Goldfeder,  Rachel L. and Enns,  Gregory M. and David,  Sean P. and Pakdaman,  Neda and Ormond,  Kelly E. and Caleshu,  Colleen and Kingham,  Kerry and Klein,  Teri E. and Whirl-Carrillo,  Michelle and Sakamoto,  Kenneth and Wheeler,  Matthew T. and Butte,  Atul J. and Ford,  James M. and Boxer,  Linda and Ioannidis,  John P. A. and Yeung,  Alan C. and Altman,  Russ B. and Assimes,  Themistocles L. and Snyder,  Michael and Ashley,  Euan A. and Quertermous,  Thomas},
  year = {2014},
  month = mar,
  pages = {1035}
}

@article{Jacobsen2022,
  title = {Phenotype‐driven approaches to enhance variant prioritization and diagnosis of rare disease},
  volume = {43},
  ISSN = {1098-1004},
  url = {http://dx.doi.org/10.1002/humu.24380},
  DOI = {10.1002/humu.24380},
  number = {8},
  journal = {Human Mutation},
  publisher = {Wiley},
  author = {Jacobsen,  Julius O. B. and Kelly,  Catherine and Cipriani,  Valentina and Research Consortium,  Genomics England and Mungall,  Christopher J. and Reese,  Justin and Danis,  Daniel and Robinson,  Peter N. and Smedley,  Damian},
  year = {2022},
  month = apr,
  pages = {1071–1081}
}

@article{Robinson2020,
  title = {Interpretable Clinical Genomics with a Likelihood Ratio Paradigm},
  volume = {107},
  ISSN = {0002-9297},
  url = {http://dx.doi.org/10.1016/j.ajhg.2020.06.021},
  DOI = {10.1016/j.ajhg.2020.06.021},
  number = {3},
  journal = {The American Journal of Human Genetics},
  publisher = {Elsevier BV},
  author = {Robinson,  Peter N. and Ravanmehr,  Vida and Jacobsen,  Julius O.B. and Danis,  Daniel and Zhang,  Xingmin Aaron and Carmody,  Leigh C. and Gargano,  Michael A. and Thaxton,  Courtney L. and Karlebach,  Guy and Reese,  Justin and Holtgrewe,  Manuel and K\"{o}hler,  Sebastian and McMurry,  Julie A. and Haendel,  Melissa A. and Smedley,  Damian},
  year = {2020},
  month = sep,
  pages = {403–417}
}

@article{Jin2021,
  title = {What Disease Does This Patient Have? A Large-Scale Open Domain Question Answering Dataset from Medical Exams},
  volume = {11},
  ISSN = {2076-3417},
  url = {http://dx.doi.org/10.3390/app11146421},
  DOI = {10.3390/app11146421},
  number = {14},
  journal = {Applied Sciences},
  publisher = {MDPI AG},
  author = {Jin,  Di and Pan,  Eileen and Oufattole,  Nassim and Weng,  Wei-Hung and Fang,  Hanyi and Szolovits,  Peter},
  year = {2021},
  month = jul,
  pages = {6421}
}

@article{Wu2025,
  title = {An automated framework for assessing how well LLMs cite relevant medical references},
  volume = {16},
  ISSN = {2041-1723},
  url = {http://dx.doi.org/10.1038/s41467-025-58551-6},
  DOI = {10.1038/s41467-025-58551-6},
  number = {1},
  journal = {Nature Communications},
  publisher = {Springer Science and Business Media LLC},
  author = {Wu,  Kevin and Wu,  Eric and Wei,  Kevin and Zhang,  Angela and Casasola,  Allison and Nguyen,  Teresa and Riantawan,  Sith and Shi,  Patricia and Ho,  Daniel and Zou,  James},
  year = {2025},
  month = apr 
}

@article{Asgari2025,
  title = {A framework to assess clinical safety and hallucination rates of LLMs for medical text summarisation},
  volume = {8},
  ISSN = {2398-6352},
  url = {http://dx.doi.org/10.1038/s41746-025-01670-7},
  DOI = {10.1038/s41746-025-01670-7},
  number = {1},
  journal = {npj Digital Medicine},
  publisher = {Springer Science and Business Media LLC},
  author = {Asgari,  Elham and Montaña-Brown,  Nina and Dubois,  Magda and Khalil,  Saleh and Balloch,  Jasmine and Yeung,  Joshua Au and Pimenta,  Dominic},
  year = {2025},
  month = may 
}

@article{Mao2024,
  title = {AI-MARRVEL — A Knowledge-Driven AI System for Diagnosing Mendelian Disorders},
  volume = {1},
  ISSN = {2836-9386},
  url = {http://dx.doi.org/10.1056/AIoa2300009},
  DOI = {10.1056/aioa2300009},
  number = {5},
  journal = {NEJM AI},
  publisher = {Massachusetts Medical Society},
  author = {Mao,  Dongxue and Liu,  Chaozhong and Wang,  Linhua and AI-Ouran,  Rami and Deisseroth,  Cole and Pasupuleti,  Sasidhar and Kim,  Seon Young and Li,  Lucian and Rosenfeld,  Jill A. and Meng,  Linyan and Burrage,  Lindsay C. and Wangler,  Michael F. and Yamamoto,  Shinya and Santana,  Michael and Perez,  Victor and Shukla,  Priyank and Eng,  Christine M. and Lee,  Brendan and Yuan,  Bo and Xia,  Fan and Bellen,  Hugo J. and Liu,  Pengfei and Liu,  Zhandong},
  year = {2024},
  month = apr 
}

@article{Tideman1987,
  title = {Independence of clones as a criterion for voting rules},
  volume = {4},
  ISSN = {1432-217X},
  url = {http://dx.doi.org/10.1007/BF00433944},
  DOI = {10.1007/bf00433944},
  number = {3},
  journal = {Social Choice and Welfare},
  publisher = {Springer Science and Business Media LLC},
  author = {Tideman,  T. N.},
  year = {1987},
  month = sep,
  pages = {185–206}
}

@article{Richards2015,
  title = {Standards and guidelines for the interpretation of sequence variants: a joint consensus recommendation of the American College of Medical Genetics and Genomics and the Association for Molecular Pathology},
  volume = {17},
  ISSN = {1098-3600},
  url = {http://dx.doi.org/10.1038/gim.2015.30},
  DOI = {10.1038/gim.2015.30},
  number = {5},
  journal = {Genetics in Medicine},
  publisher = {Elsevier BV},
  author = {Richards,  Sue and Aziz,  Nazneen and Bale,  Sherri and Bick,  David and Das,  Soma and Gastier-Foster,  Julie and Grody,  Wayne W. and Hegde,  Madhuri and Lyon,  Elaine and Spector,  Elaine and Voelkerding,  Karl and Rehm,  Heidi L.},
  year = {2015},
  month = may,
  pages = {405–424}
}

@misc{ClinGenSVIGuidance2026,
  author       = {{Clinical Genome Resource (ClinGen)}},
  title        = {ClinGen Variant Classification Guidance},
  howpublished = {\url{https://www.clinicalgenome.org/tools/clingen-variant-classification-guidance/}},
  year         = {2026},
  note         = {Accessed 2026-02-25}
}

@article{Oza2018HLVCEP,
  title = {Expert specification of the {ACMG}/{AMP} variant interpretation guidelines for genetic hearing loss},
  volume = {39},
  ISSN = {1059-7794},
  url = {https://doi.org/10.1002/humu.23630},
  DOI = {10.1002/humu.23630},
  number = {11},
  journal = {Human Mutation},
  publisher = {Wiley},
  author = {Oza, Andrea M. and DiStefano, Marina T. and Hemphill, Sarah E. and others},
  year = {2018},
  month = nov,
  pages = {1593--1613}
}

@article{Kelly2018MYH7VCEP,
  title = {Adaptation and validation of the {ACMG}/{AMP} variant classification framework for {MYH7}-associated inherited cardiomyopathies: recommendations by ClinGen's inherited cardiomyopathy expert panel},
  volume = {20},
  ISSN = {1098-3600},
  url = {https://doi.org/10.1038/gim.2017.218},
  DOI = {10.1038/gim.2017.218},
  number = {3},
  journal = {Genetics in Medicine},
  publisher = {Elsevier BV},
  author = {Kelly, Melissa A. and Caleshu, Colleen and Morales, Ana and others},
  year = {2018},
  month = mar,
  pages = {351--359}
}

@misc{ValsMedQA2025,
  author       = {{Vals AI}},
  title        = {MedQA: Benchmark Results},
  howpublished = {\url{https://www.vals.ai/benchmarks/medqa}},
  year         = {2025},
  month        = oct,
  note         = {Accessed 2025-10-30}
}

@inproceedings{WangSelfCons2023,
title	= {Self-Consistency Improves Chain of Thought Reasoning in Language Models},
author	= {Xuezhi Wang and Jason Wei and Dale Schuurmans and Quoc V. Le and Ed H. Chi and Sharan Narang and Aakanksha Chowdhery and Denny Zhou},
year	= {2023},
URL	= {https://arxiv.org/abs/2203.11171},
booktitle	= {ICLR 2023}
}

@inproceedings{YaoTreeThoughts2023,
author = {Yao, Shunyu and Yu, Dian and Zhao, Jeffrey and Shafran, Izhak and Griffiths, Thomas L. and Cao, Yuan and Narasimhan, Karthik},
title = {Tree of thoughts: deliberate problem solving with large language models},
year = {2023},
publisher = {Curran Associates Inc.},
address = {Red Hook, NY, USA},
booktitle = {Proceedings of the 37th International Conference on Neural Information Processing Systems},
articleno = {517},
numpages = {14},
location = {New Orleans, LA, USA},
series = {NIPS '23}
}

@inproceedings{WangMathShepherd2024,
  title = {Math-Shepherd: Verify and Reinforce LLMs Step-by-step without Human Annotations},
  url = {http://dx.doi.org/10.18653/v1/2024.acl-long.510},
  DOI = {10.18653/v1/2024.acl-long.510},
  booktitle = {Proceedings of the 62nd Annual Meeting of the Association for Computational Linguistics (Volume 1: Long Papers)},
  publisher = {Association for Computational Linguistics},
  author = {Wang,  Peiyi and Li,  Lei and Shao,  Zhihong and Xu,  Runxin and Dai,  Damai and Li,  Yifei and Chen,  Deli and Wu,  Yu and Sui,  Zhifang},
  year = {2024},
  pages = {9426–9439}
}

@inproceedings{Chaing2021,
  author = {Chiang, Chun-Wei and Yin, Ming},
  title = {You'd Better Stop! Understanding Human Reliance on Machine Learning Models under Covariate Shift},
  booktitle = {Proceedings of the 13th ACM Web Science Conference 2021},
  year = {2021},
  pages = {120--129},
  doi = {10.1145/3447535.3462487},
  url = {https://dl.acm.org/doi/fullHtml/10.1145/3447535.3462487},
  publisher = {Association for Computing Machinery},
  address = {New York, NY, USA}
}

@inproceedings{Yin2019,
  author = {Yin, Ming and Vaughan, Jennifer Wortman and Wallach, Hanna},
  title = {Understanding the Effect of Accuracy on Trust in Machine Learning Models},
  booktitle = {Proceedings of the 2019 CHI Conference on Human Factors in Computing Systems},
  year = {2019},
  articleno = {279},
  pages = {1--12},
  doi = {10.1145/3290605.3300509},
  url = {https://doi.org/10.1145/3290605.3300509},
  publisher = {Association for Computing Machinery},
  address = {New York, NY, USA}
}

@article{LeeConSol2025,
  title={ConSol: Sequential Probability Ratio Testing to Find Consistent LLM Reasoning Paths Efficiently},
  author={Lee, Jaeyeon and Qi, Guantong and Neeley, Matthew Brady and Liu, Zhandong and Jeong, Hyun-Hwan},
  journal={arXiv preprint arXiv:2503.17587},
  year={2025}
}

@article{Murdock2020,
  title = {Enhancing Diagnosis Through RNA Sequencing},
  volume = {40},
  ISSN = {0272-2712},
  url = {http://dx.doi.org/10.1016/j.cll.2020.02.001},
  DOI = {10.1016/j.cll.2020.02.001},
  number = {2},
  journal = {Clinics in Laboratory Medicine},
  publisher = {Elsevier BV},
  author = {Murdock,  David R.},
  year = {2020},
  month = jun,
  pages = {113–119}
}

@book{IOM1994GeneticRisks,
  author    = {Institute of Medicine (US) Committee on Assessing Genetic Risks and Andrews, Lori B. and Fullarton, Jane E. and Holtzman, Neil A. and others},
  editor    = {Andrews, Lori B. and Fullarton, Jane E. and Holtzman, Neil A.},
  title     = {Assessing Genetic Risks: Implications for Health and Social Policy},
  publisher = {National Academies Press (US)},
  address   = {Washington, DC},
  year      = {1994},
  chapter   = {2, Genetic Testing and Assessment},
  url       = {https://www.ncbi.nlm.nih.gov/books/NBK236037/},
  note      = {Available from: National Center for Biotechnology Information (NCBI) Bookshelf}
}

@article{Goddard2012AutomationBias,
  author = {Goddard, K. and Roudsari, A. and Wyatt, J. C.},
  title = {Automation bias: a systematic review of frequency, effect mediators, and mitigators},
  journal = {Journal of the American Medical Informatics Association},
  year = {2012},
  month = jan,
  volume = {19},
  number = {1},
  pages = {121--127},
  doi = {10.1136/amiajnl-2011-000089},
  pmid = {21685142},
  pmcid = {PMC3240751},
  url = {https://pubmed.ncbi.nlm.nih.gov/21685142/}
}

@article{Zhao2026DeepRare,
  title = {An agentic system for rare disease diagnosis with traceable reasoning},
  author = {Zhao, Xinyu and Team, DeepRare and others},
  journal = {Nature},
  year = {2026},
  month = feb,
  DOI = {10.1038/s41586-025-10097-9},
  url = {https://doi.org/10.1038/s41586-025-10097-9},
  note = {Published online 2026-02-18}
}

@article{Passi2022OverrelianceAI,
  title = {Overreliance on AI: Literature review},
  author = {Passi, Samir and Vorvoreanu, Mihaela},
  institution = {Microsoft Research},
  year = {2022},
  month = jun,
  url = {https://www.microsoft.com/en-us/research/wp-content/uploads/2022/06/Aether-Overreliance-on-AI-Review-Final-6.21.22.pdf},
  note = {Aether Committee publication}
}

@article{Novak2024,
  author = {Novak, A. and O'Neill, J. and Wilson, C. and Flannigan, C. and Frean, M. and Lim, C. and Body, R. and Gerry, S.},
  title = {Can AI improve chest radiograph interpretation for suspected pneumothorax? A multicentre, multireader, multicase, randomised trial},
  journal = {Emergency Medicine Journal},
  year = {2024},
  volume = {41},
  number = {10},
  pages = {601--608},
  doi = {10.1136/emermed-2023-213921},
  pmid = {39009424},
  url = {https://pubmed.ncbi.nlm.nih.gov/39009424/}
}

@article{Lyell2025,
  author = {Lyell, D. and others},
  title = {The impact of AI support on emergency doctors' diagnosis and management: a multicentre, multireader, multicase crossover randomised controlled trial},
  journal = {Emergency Medicine Journal},
  year = {2025},
  doi = {10.1136/emermed-2024-214925},
  pmid = {41083204},
  url = {https://pubmed.ncbi.nlm.nih.gov/41083204/}
}

@article{Alsentzer2025SHEPHERD,
  author = {Alsentzer, Emily and others},
  title = {Few shot learning for phenotype-driven diagnosis of patients with rare genetic diseases},
  journal = {npj Digital Medicine},
  year = {2025},
  volume = {8},
  pages = {157},
  doi = {10.1038/s41746-025-01749-1},
  url = {https://www.nature.com/articles/s41746-025-01749-1}
}

@article{Chen2025MAC,
  author = {Chen, Yan and others},
  title = {Enhancing diagnostic capability with multi-agents conversational large language models},
  journal = {npj Digital Medicine},
  year = {2025},
  volume = {8},
  pages = {54},
  doi = {10.1038/s41746-025-01550-0},
  url = {https://www.nature.com/articles/s41746-025-01550-0}
}

@online{HGNC_VARS2,
  title   = {VARS2},
  author  = {{HGNC: HUGO Gene Nomenclature Committee}},
  url     = {https://www.genenames.org/data/gene-symbol-report/#!/symbol/VARS2},
  urldate = {2026-02-12},
}

@online{HGNC_ANO5,
  title   = {ANO5},
  author  = {{HGNC: HUGO Gene Nomenclature Committee}},
  url     = {https://www.genenames.org/data/gene-symbol-report/#!/symbol/ANO5},
  urldate = {2026-02-12},
}

@online{HGNC_CHD8,
  title   = {CHD8},
  author  = {{HGNC: HUGO Gene Nomenclature Committee}},
  url     = {https://www.genenames.org/data/gene-symbol-report/#!/symbol/CHD8},
  urldate = {2026-02-12},
}

@online{HGNC_AP4M1,
  title   = {AP4M1},
  author  = {{HGNC: HUGO Gene Nomenclature Committee}},
  url     = {https://www.genenames.org/data/gene-symbol-report/#!/symbol/AP4M1},
  urldate = {2026-02-12},
}

@online{AWS_CLAUDEPARAMS,
  title        = {Model parameters for Anthropic Claude messages inference requests},
  author       = {{Amazon Web Services}},
  year         = {2025},
  url          = {https://docs.aws.amazon.com/bedrock/latest/userguide/model-parameters-anthropic-claude-messages-request-response.html},
  note         = {Amazon Bedrock User Guide},
  urldate      = {2025-11-04}
}

@online{AWS_HIPAA_ELIGIBLE_SERVICES,
  title        = {HIPAA Eligible Services Reference},
  author       = {{Amazon Web Services}},
  year         = {2026a},
  url          = {https://aws.amazon.com/compliance/hipaa-eligible-services-reference/},
  note         = {AWS Compliance},
  urldate      = {2026-02-26}
}

@online{AWS_BEDROCK_DATAPROTECTION,
  title        = {Data protection in Amazon Bedrock},
  author       = {{Amazon Web Services}},
  year         = {2026b},
  url          = {https://docs.aws.amazon.com/bedrock/latest/userguide/data-protection.html},
  note         = {Amazon Bedrock User Guide},
  urldate      = {2026-02-26}
}

@online{AWS_BEDROCK_INVOCATIONLOGGING,
  title        = {Monitor Amazon Bedrock using invocation logging},
  author       = {{Amazon Web Services}},
  year         = {2026c},
  url          = {https://docs.aws.amazon.com/bedrock/latest/userguide/model-invocation-logging.html},
  note         = {Amazon Bedrock User Guide},
  urldate      = {2026-02-26}
}
\end{footnotesize}

\newpage

\begin{algorithm}[H]
\caption{TidemanRankedPairs(\textit{ballots})}
\label{alg:rankedpairs}
\begin{algorithmic}[1]
\Require \textit{ballots}: list of partial rankings over candidates (1 = best). Missing rank = unranked.
\Ensure \textit{order}: candidates from best to worst
\State \textbf{if } ballots is empty \textbf{ then } \Return empty list
\State $C \gets$ sorted set of all candidates appearing in any ballot
\State $m \gets |C|$

\Statex
\State \Comment{\textbf{Borda scores for deterministic tie-breaks}}
\State initialize $\text{Borda}[c] \gets 0$ for all $c \in C$
\For{each ballot $s$}
  \State let $r$ be $s$ reindexed to $C$ (candidates absent in $s$ are unranked in $r$)
  \For{each $c \in C$}
    \If{$r[c]$ is ranked}
      \State $\text{Borda}[c] \gets \text{Borda}[c] + (m - r[c] + 1)$
    \EndIf
  \EndFor
\EndFor

\Statex
\State \Comment{\textbf{Pairwise wins (unranked treated as worse than any ranked)}}
\State initialize $\text{Wins}[a][b] \gets 0$ for all $a,b \in C$
\For{each ballot $s$}
  \State let $r$ be $s$ reindexed to $C$
  \For{each ordered pair $(a,b)$ with $a \neq b$}
    \If{$r[a]$ ranked and $r[b]$ unranked}
      \State $\text{Wins}[a][b] \gets \text{Wins}[a][b] + 1$
    \ElsIf{$r[a]$ ranked and $r[b]$ ranked \textbf{and} $r[a] < r[b]$}
      \State $\text{Wins}[a][b] \gets \text{Wins}[a][b] + 1$
    \EndIf
    \Comment{(If $a$ unranked and $b$ ranked, do nothing)}
  \EndFor
\EndFor

\Statex
\State \Comment{\textbf{Compute strength-of-victory pairs and sort}}
\State $P \gets$ empty list of directed pairs
\For{each unordered pair $\{a,b\}$ with $a \neq b$ and $a < b$ (lexicographic)}
  \State $w_{ab} \gets \text{Wins}[a][b]$; $w_{ba} \gets \text{Wins}[b][a]$
  \If{$w_{ab} > w_{ba}$}
    \State append $(a,b,\; w_{ab} - w_{ba},\; w_{ab},\; w_{ba})$ to $P$
  \ElsIf{$w_{ba} > w_{ab}$}
    \State append $(b,a,\; w_{ba} - w_{ab},\; w_{ba},\; w_{ab})$ to $P$
  \EndIf
\EndFor
\State sort $P$ in descending order by:
\Statex \hspace{1.8em} (i) victory margin, (ii) winner's votes, then ascending by (iii) winner name, (iv) loser name

\Statex
\State \Comment{\textbf{Lock pairs without creating cycles}}
\State initialize directed adjacency sets $\text{Adj}[c] \gets \emptyset$ for all $c \in C$
\Function{CreatesCycle}{$u, v$}
  \State \textbf{return} (there exists a path $v \leadsto u$ in graph $\text{Adj}$)
  \Comment{e.g., DFS from $v$ following $\text{Adj}$ edges}
\EndFunction
\For{each $(u,v, \cdot)$ in $P$ in order}
  \If{\textbf{not} \Call{CreatesCycle}{$u,v$}}
    \State add edge $u \to v$ to $\text{Adj}$
  \EndIf
\EndFor

\Statex
\State \Comment{\textbf{Topological ordering with tie-breaks (Borda, then name)}}
\State compute $\text{InDeg}[c]$ from $\text{Adj}$
\State $\text{Avail} \gets$ candidates with $\text{InDeg}=0$, sorted by descending $\text{Borda}$, then by name
\State $\text{Order} \gets$ empty list
\While{$\text{Avail}$ not empty}
  \State remove first $u$ from $\text{Avail}$; append $u$ to $\text{Order}$
  \For{each $v$ in $\text{Adj}[u]$ (any order)}
    \State $\text{InDeg}[v] \gets \text{InDeg}[v] - 1$
    \If{$\text{InDeg}[v] = 0$}
      \State insert $v$ into $\text{Avail}$
    \EndIf
  \EndFor
  \State resort $\text{Avail}$ by descending $\text{Borda}$, then by name
\EndWhile

\State \Return $\text{Order}$
\end{algorithmic}
\end{algorithm}

\begin{algorithm}[H]
\caption{ReorderListByConsensus(\textit{PrioritizedLists}, \label{alg:reorder}
\textit{InitialList})}
\begin{algorithmic}[1]
\Require \textit{PrioritizedLists}: outputs of LLM; \textit{InitialList}: Top-G genes ordered by the score of AI-MARRVEL
\Ensure \textit{ReorderedList}: reranked gene list in consensus order, then any stragglers

\Statex \Comment{\textbf{Build Ballots By Completing the Prioritized Lists with InitialList}}
\State $\text{ballots} \gets [\ ]$ \Comment{each ballot is a map: gene $\to$ rank (0-based)}
\For{$\text{PrioritizedList} \in \textit{PrioritizedLists}$}
    \State $\text{Head} \gets$ items of \textit{InitialList} which is in \text{PrioritizedList}, ordered by \text{PrioritizedList}
    \State $\text{Tail} \gets$ items of \textit{InitialList} which is not in \text{PrioritizedList} (ordered by InitialList)
    \State $\text{ballot} \gets$ concatenate \text{Head} then \text{Tail}

  \State append $\text{ballot}$ to $\text{ballots}$
\EndFor
\Statex

\State $\text{ConsensusOrder} \gets$ \Call{TidemanRankedPairs}{ballots}
\State $\text{InConsensus} \gets$ subsequence of \textit{ConsensusOrder} that appear in \textit{InitialList}
\State $\text{Head} \gets$ items of \textit{InitialList} which is in \text{InConsensus}, ordered by \text{InConsensus}
\State $\text{Tail} \gets$ items of \textit{InitialList} which is not in \text{InConsensus} (ordered by InitialList)
\State $\text{ReorderedList} \gets$ concatenate \text{Head} then \text{Tail}
\Statex

\State \Return $\text{ReorderedList}$
\end{algorithmic}
\end{algorithm}

\newpage



\begin{table}
\captionsetup{margin={0.5in,0in}} 

\caption{\textcolor{Highlight}{Stats of BG-DDD-UDN Cohort Data}}
\centering
\begin{small}

\begin{tabular}{lccc}
\toprule
 & \textbf{BG} & \textbf{DDD} & \textbf{UDN} \\
\midrule
Case & 63 & 214 & 90 \\
Avg. Gene & 1469.2 & 811.5 & 1072.2 \\
Avg. HPO & 10.59 & 7.38 & 43.98 \\
Avg. Causal Genes & 1.175 & 1.000 & 1.044 \\
Unique Causal Genes & 74 & 116 & 93 \\
Accessibility & Internal & Restricted & Restricted \\
\midrule
Source & Baylor Genetics & DDD & Undiagnosed Diseases Network \\
Curation & In-House & AMELIE & In-House \\
\bottomrule
\end{tabular}

\end{small}
\label{tab:1}
\end{table}

\begin{table}[]
    \captionsetup{margin={0.5in,0in}} 
    
    \caption{\textcolor{Highlight}{HPO summary for nystagmus}}
    \footnotesize
    \centering
    \begin{tabularx}{\textwidth}{
      >{\raggedright\arraybackslash}p{2cm}
      >{\raggedright\arraybackslash}p{3cm}
      >{\raggedright\arraybackslash}X
      >{\raggedright\arraybackslash}X}
    \toprule
    ID & Name & Synonyms & Definition \\
    \midrule
    HP:0000639 & Nystagmus &
    ['Involuntary, rapid, rhythmic eye movements'] &
    Rhythmic, involuntary oscillations of one or both eyes related to abnormality in fixation, conjugate gaze, or vestibular mechanisms. \\
    \bottomrule
    \end{tabularx}
    \label{tab:2}
\end{table}

\begin{table}[h]
    \captionsetup{margin={0.5in,0in}} 

    \caption{\textcolor{Highlight}{Gene summary for \textit{SPG7}}}
    \footnotesize
    \centering
    \begin{tabularx}{\textwidth}{
      >{\raggedright\arraybackslash}p{.5cm}  
      >{\raggedright\arraybackslash}p{2cm}  
      >{\raggedright\arraybackslash}p{1cm}  
      >{\raggedright\arraybackslash}p{1cm}  
      >{\raggedright\arraybackslash}p{1.5 cm}  
      >{\raggedright\arraybackslash}p{1cm}  
      >{\raggedright\arraybackslash}p{1.5cm}  
      >{\raggedright\arraybackslash}p{1.5cm}  
      >{\raggedright\arraybackslash}X}        
    \toprule
    Gene & ClinVar status & CADD Score & AF & Zygosity & Trans-Heterozygous & Consequence & Impact & Phenotype \\
    \midrule
    SPG7 &
    Conflicting interpretations of pathogenicity &
    27.8 &
    0.002923 &
    Heterozygous &
    No &
    missense variant &
    MODERATE &
    \textbf{Spastic paraplegia 7, autosomal recessive}. Inheritance: AD; AR. Clinical features:  Nystagmus; Lower limb muscle weakness; Urinary bladder sphincter dysfunction; Waddling gait; Lower limb hypertonia; Muscle weakness; Degeneration of the lateral corticospinal tracts; Hyperreflexia; Vertical supranuclear gaze palsy; Dysphagia; Scoliosis; Babinski sign; Gait disturbance; Memory impairment; Pes cavus; Spastic ataxia; Urinary urgency; Lower limb hyperreflexia; Upper limb muscle weakness; Dysarthria; Dysdiadochokinesis; Upper limb hypertonia; Postural instability; Limb ataxia; Upper limb hyperreflexia; Supranuclear gaze palsy; Abnormality of somatosensory evoked potentials; Optic atrophy; Spastic paraplegia; Upper limb spasticity. \\
    \bottomrule
    \end{tabularx}
    \label{tab:3}
\end{table}

\newpage

\end{document}